\documentclass[twocolumn,floatfix,
nofootinbib,
prd,aps,superscriptaddress,tightenlines]{revtex4}
\usepackage{epsfig}
\usepackage{pstricks}

\newcommand{\xslash}[1]{{\rlap{$#1$}/}}

\def\OMIT#1{{}}
\def\lqcd{\Lambda_{\rm QCD}}

\newcommand{\nn}{\nonumber \\ }

\newcommand{\me}[3]{\ensuremath{\left\langle{#1}\vphantom{#2 #3}
\right|{#2}\left|\vphantom{#1 #2}{#3}\right\rangle}}

\newcommand{\bn}{{\bar n}}

\def\abs#1{ \left| #1 \right| }
\def\ket#1{ \left| #1 \right\rangle }
\def\bra#1{ \left\langle #1 \right| }
\def\darr#1{\raise1.5ex\hbox{$\leftrightarrow$}\mkern-16.5mu #1}

\def\diff{{\text d}}

\def\euv{ \epsilon_{\text{UV}} }
\def\eir{ \epsilon_{\text{IR}} }

\def\bN{\bar N}

\begin{document}

\title{Deep inelastic scattering as $x \to 1$ using soft-collinear effective theory}

\author{Aneesh V.~Manohar}
\affiliation{Department of Physics, University of California at San Diego,
  La Jolla, CA 92093\vspace{4pt} }

\date{September 2003}

\begin{abstract}
Soft-collinear effective theory (SCET) is used to sum Sudakov double-logarithms in the $x\to1$ endpoint region for the deep inelastic scattering structure function. The calculations are done in both the target rest frame and the Breit frame. The separation of scales in the effective theory implies that the anomalous dimension of the SCET current is  linear in $\ln \mu$, and the anomalous dimension for the $N^{\text{th}}$ moment of the structure function is linear in $\ln N$, to all orders in perturbation theory. The SCET formulation is shown to be free of Landau pole singularities. Some important differences between the deep inelastic structure function and the shape function in $B$ decay are discussed. 
\end{abstract}

\maketitle

\section{Introduction}

The deep inelastic scattering cross-section is the inclusive cross-section for lepton scattering off a hadronic target at large momentum transfer. The cross-section is conventionally written in terms of structure functions of the momentum transfer $Q^2$ and a dimensionless variable $0 \le x \le 1$. The structure function(s) 
$F(x,Q^2)$ cannot be computed in perturbation theory, but its $Q^2$ dependence can. The structure function contains large logarithms of the form $\left( \alpha_s \ln Q^2/\lqcd^2 \right)^n$, which can be summed by evolving $F(x,Q^2)$ from the large scale $Q^2$ to a lower scale $\mu$ using the renormalization group equations. $\mu$ is chosen to be  a few GeV, parametrically of the order of $\lqcd$, but still large enough that perturbation theory is valid.

As $x \to 1$, there are additional large logarithms that need to be summed to get a reliable evaluation of the scattering cross-section. The invariant mass of the final hadronic state is
\begin{eqnarray}
M^2_X &=& {Q^2(1-x) \over x}\ ,
\label{1.01}
\end{eqnarray}
and $M^2_X \to 0$ as $x \to 1$. The total cross-section is infrared finite, even though the real and virtual emission processes are separately infrared divergent. In the region $x \to 1$, real gluon emission is suppressed, and the cancellation between real and virtual emission becomes more delicate, leading to large corrections to the cross-section. The form of the perturbation series is most conveniently described in moment space, where $x \to 1$ corresponds to large moments, $N \to \infty$, with the heuristic rule $1-x \sim 1/N$. As $N \to \infty$, the structure function moments contain terms of the form $\alpha_s^r \ln^s N$ with $ s \le 2 r$. These Sudakov double-logarithms are important in the endpoint region. The summation of these terms is well-known, and has been discussed extensively in the literature~\cite{tasi}. The general result is that the $N^{\text{th}}$ moment of the structure function at $Q^2$, $F_N(Q^2)$ can be written as~\cite{Catani,tasi}
\begin{eqnarray}
\ln F_N(Q^2) &=&  f_0 \left( \alpha_s \ln N \right) \ln N +f_1 \left( \alpha_s \ln N \right) \nn
&& + \alpha_s f_2 \left( \alpha_s \ln N \right) + \ldots\ \ .
\label{1.02}
\end{eqnarray}
The exponential of $ f_0 \left( \alpha_s \ln N \right) \ln N$ gives the leading Sudakov double-logarithmic series.

In this paper, Sudakov double logarithms in the endpoint region are calculated using soft-collinear effective theory (SCET)~\cite{BFL,SCET}.  SCET allows one to compute the cross-section in the endpoint region in a systematic expansion to any desired accuracy. The results are free of Landau pole singularities. The calculations are described in detail in both the target rest frame and the Breit frame, and give an instructive example of the use of SCET. There are several unusual aspects of SCET which are discussed here; the frame dependence of the way in which infrared divergences cancel between the soft and collinear modes,  and the relation between structure functions and local operators which is different for deep inelastic scattering and  $B$ decays. Consistency of the effective theory implies that the SCET anomalous dimension is linear in $\ln \mu$, which leads to the form Eq.~(\ref{1.02}) of the perturbation series. The SCET calculation to the accuracy presented here gives the moments of the structure function including the first two exponentiated series $f_{0,1}$ in Eq.~(\ref{1.02}), as well as all terms of order $\alpha_s$ which  do not vanish as $N \to \infty$. The method used parallels that for $B \to X_s \gamma$ in Ref.~\cite{BFL}. Deep inelastic scattering structure functions in the Breit frame (but not in the endpoint region) were considered in Ref.~\cite{Ira}. The SCET formalism used in this paper is described in Refs.~\cite{SCET}.

\section{Outline of Calculation}

The calculation of the deep inelastic scattering cross-section will be performed using a sequence of effective field theories. The scattering amplitude involves the interaction of a lepton beam with a hadron target via a virtual photon. The leptonic interactions are calculable using QED, and will not be discussed here. The quantity of interest is the interaction of the virtual photon with the hadronic target.

\begin{figure}
\includegraphics[width=10cm]{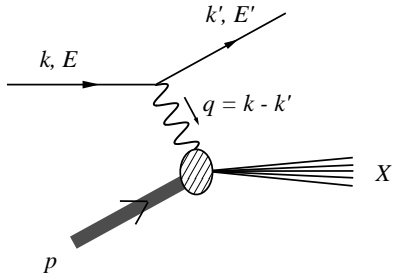}
\caption{The deep inelastic scattering process. The incoming lepton with energy $E$ scatters off a hadronic target with momentum $p$ to produce the final hadronic state $X$. \label{fig:dis}}
\end{figure}

At scales much larger than $Q^2$, the interaction of photons with hadrons is described using the electromagnetic current of quarks interacting via the full QCD Lagrangian. The hadronic scattering amplitude is the matrix element of the electromagnetic current between the initial and final hadronic states  (see Fig.~\ref{fig:dis}),
\begin{eqnarray}
A^\mu &=& \me{X}{j^\mu}{P} \ .
\label{3.01}
\end{eqnarray}
The cross-section involves the square of the scattering amplitude, and can be written in terms of the product of two currents, summed over intermediate states,
\begin{eqnarray}
\sum_X \me{P}{j^\mu}{X} \me{X}{j^\nu}{P}\ .
\label{3.02}
\end{eqnarray}
The invariant mass of the final hadronic state for generic values of $x$ is of order $Q^2$. The final hadronic state $X$ can be integrated out at the scale $Q^2$, and the product of currents in Eq.~(\ref{3.02}) is replaced by a sum over local twist-two operators when one takes moments of the cross-section. This is the conventional method of computing the deep inelastic scattering cross-section.

As $x \to 1$, the invariant mass of the final hadronic state tends to zero.  The invariant mass of the final hadronic state is taken to be of order $Q^2 \lambda$, with $Q^2 \gg Q^2 \lambda \gg \lqcd^2$, and all results are computed in an expansion in $\lambda$. More details about this power counting scheme are given in Sec.~\ref{ssec:power}. The scale $Q^2 \lambda$ is an infrared scale for the theory at $Q$, and can be set to zero at leading order in an expansion in $\lambda$. In this limit, the final hadronic state is massless, and cannot be integrated out. Instead, the final hadronic state can be treated as a massless light-like jet, and is described by a collinear quark in SCET. Thus for $x \to 1$, the QCD current at scale $Q$ is matched onto an SCET current at scale $Q$. The coordinate axes are chosen so that the outgoing quark travels in the $n^\mu=(1,0,0,1)$ direction, and is described by the collinear field $\xi_n$ of SCET. In the target rest frame, the incoming quark has momentum of order the hadron target momentum $\lqcd$, and is an ultrasoft quark. In the Breit frame, the struck quark is back-scattered, so the incoming quark is a $\bn$ collinear quark, and is described by the field $\xi_{\bn}$, where $\bn=(1,0,0,-1)$. The first step of the calculation is the matching of the electromagnetic current  $\bar \psi \gamma^\mu \psi$ onto the SCET current $\bar \xi_n \gamma^\mu \psi$ or $\bar \xi_n \gamma^\mu \xi_{\bn}$ in the target or Breit frame, respectively. The matching coefficient is evaluated in Sec.~\ref{sec:matchQ}.

The next scale in the problem is the invariant mass of the hadronic state, $p_X^2 \sim Q^2 \lambda \sim Q^2(1-x)$. The SCET current is run from $\mu=Q$ to this scale using the  anomalous dimension, which is computed in Sec.~\ref{sec:Qrun} in both the target and Breit frames. 

At the scale $Q^2(1-x) \sim Q^2 \lambda$, $p_X^2$ is treated as large, and so the final state can be integrated out. The time-ordered product of two SCET currents can be replaced by a bilocal light-cone operator whose target matrix element is the parton distribution function. This is done by integrating out the $\xi_n$ field in SCET; the resulting operator is written in terms of ultrasoft quark fields $\psi_u$ in the target frame, and in terms of $\bn$-collinear quark fields in the Breit frame. The bilocal operator is closely related to the Collins-Soper operator~\cite{Collins}. The matching at scale $Q^2 \lambda$ is discussed in Sec.~\ref{sec:matchQl}

The last step, discussed in Sec.~\ref{sec:run}, is to run the bilocal operators from $Q^2(1-x)$ to some low scale $\mu$, and match onto local twist-two operators which give the moment sum rules for the deep inelastic structure functions.

The computations in this paper are given in Feynman gauge. The results, however, are gauge invariant, and valid in any gauge. In the effective theory, one has separate gauge invariance for the ultrasoft, $n$-collinear and $\bn$-collinear gluons. This has been checked by explicit computation.
 
The entire analysis is presented for QCD with a single quark flavor of unit charge, to avoid unnecessary indices. The final answer is given by summing the results of this paper over all flavors weighted with the square of their electromagnetic charges. 
In addition to the currents $\bar \xi_n \gamma^\mu \psi$ and $\bar \xi_n \gamma^\mu \xi_{\bn}$, one also has the hermitian conjugate currents $\bar \psi \gamma^\mu \xi_n$ and $\bar \xi_{\bar n} \gamma^\mu \xi_n$, which give the antiquark contribution, or equivalently, the crossed-graph contributions. By charge conjugation invariance, the matching coefficients and anomalous dimensions are the same for quarks and antiquarks. The final result thus has a sum over both quark and antiquark distributions.

\section{Kinematics\label{sec:kin}}

The scattering process is $e^- + p \to e^- + X$. The proton momentum is $P$, the incoming momentum of the virtual photon is $q$, the momentum transfer is $Q^2 = - q^2 \gg 0$, and $x$ is defined by
\begin{eqnarray}
x = - {q^2 \over 2 P \cdot q} = {Q^2 \over 2 P \cdot q}\ .
\label{2.01}
\end{eqnarray}
The coordinate axes are chosen so that the virtual photon is in the $z$ direction. It is useful to introduce the null vectors $n^\mu = (1,0,0,1)$ and $\bn^\mu=(1,0,0,-1)$ which point in the $\hat{\mathbf {z}}$ and $-\hat{\mathbf {z}}$ direction, respectively.
They satisfy the relations $n^2=0$, $\bn^2=0$, $\bn \cdot n = 2$. Any four-vector $a^\mu$ can be written as
\begin{eqnarray}
a^\mu &=& \frac12 a^+ \bn^\mu+\frac12 a^-  n^\mu + \mathbf{a}_\perp^\mu ,
\label{2.02}
\end{eqnarray}
where
\begin{eqnarray}
a^+ \equiv n \cdot a \ ,\qquad a^- \equiv \bn \cdot a\ ,
\label{2.03}
\end{eqnarray}
and $\mathbf{a}_\perp$ is in the $x-y$ plane. The dot product of two four-vectors is
\begin{eqnarray}
a \cdot b &=& \frac 1 2 a^+ b^- + \frac 1 2 a^- b^+ - \mathbf{a}_\perp \cdot \mathbf{b}_\perp\ ,
\label{2.04}
\end{eqnarray}
and the integration measure can be written in light-cone coordinates as
\begin{eqnarray}
{\rm d}^d k &=& {1\over 2}\, {\rm d}k^+\ {\rm d}k^- \ {\rm d}^{d-2} k_\perp\ .
\label{2.05}
\end{eqnarray}

\subsection{Target Rest Frame}

In the target rest frame, $\mathbf{q}_\perp=0$, and $Q^2=-q^2=-q^+ q^-$. The deep inelastic limit $Q^2 \to \infty$ with $x$ fixed is given by taking $q^- \to \infty$ at fixed $q^+$, so that $q^- \gg q^+$. Then
\begin{eqnarray}
x &=& - {q^+ q^- \over P^+ q^- + P^- q^+} \approx  -{q^+ \over P^+}\ ,
\label{2.06}
\end{eqnarray}
where $P^+=M_T$ is the target mass, so that
\begin{eqnarray}
q^+ &=& - x P^+ \ ,\nn
q^- &=& {Q^2 \over x P^+} \ ,\nn
p_X^+ &=& P^+\left(1-x\right) \ , \nn
p_X^- &=& P^- + q^-  \sim q^- \ ,\nn
p_X^2 &=& Q^2 {1 - x \over x}\ ,
\label{2.07}
\end{eqnarray}
where $p_X=P+q$ is the momentum of the final hadronic state.

\subsection{Breit Frame}

The virtual photon carries only momentum, and no energy in the Breit frame. The Breit frame is obtained from the target rest frame by boosting along the $z$-axis, so that the proton and virtual photon have no $\perp$ component of momentum in either frame. The momentum components in the Breit frame are
\begin{eqnarray}
q^+ &=& -Q, \nn
q^- &=& Q, \nn
p^+ &=& Q+l^+, \nn
p^- &=& l^- ,\nn
p_X^+ &=& l^+, \nn
p_X^- &=& Q+ l^-,
\label{2.08}
\end{eqnarray}
where $l^\pm$ are fixed by setting $P^2\approx Q l^- = M_T$ and
\begin{eqnarray}
x &=& {Q \over Q + l^+  - M_T/Q } \approx  {Q \over Q + l^+},
\label{2.09}
\end{eqnarray}
so that
\begin{eqnarray}
1-x &=& {l^+ \over Q},
\label{2.10}
\end{eqnarray}
and
\begin{eqnarray}
p_X^2 &=& Q l^+.
\label{2.11}
\end{eqnarray}

The $\pm$ components of momentum in the Breit frame are given by multiplying the $\pm$ components of momentum in the target rest frame by $\left(Q/x P^+\right)^\pm \sim \left(Q/ P^+\right)^\pm$ for $x \approx 1$.

\subsection{Power Counting\label{ssec:power}}

The endpoint region is $p_X^2 \sim Q^2(1-x) \sim Q^2 \lambda$, with $Q^2 \gg Q^2 \lambda \gg \lqcd^2$. In this region, the final state still involves a sum over many hadronic states, but has small invariant mass and is jet-like. The dimensionless power counting parameter $\lambda \ll 1$ is introduced as the expansion parameter, and $1-x \sim \lambda$. 

The power counting is simplest in the Breit frame, where $\ell^+ \sim Q \lambda$ and $\ell^- \sim \lqcd$, so that $P^+ \sim Q$, $P^- \sim \lqcd$, $p_X^+ \sim Q \lambda$, $p_X^- \sim Q$, $q^+ \sim Q$, $q^- \sim Q$. There are three important scales: (1) $Q^2$, the invariant mass of the virtual photon, (2) $Q^2 \lambda$, the invariant mass of the final state hadronic jet, and (3) $M_T^2\sim \lqcd^2$, the invariant mass of the target. The scale $Q^2 \lambda^2$ does not play an important role in deep inelastic scattering; it does for the shape function in $B$ decays~\cite{Bauer}.

Particles with $p^- \sim Q$, $p^+ \sim Q \lambda^2$ and $\mathbf{p}_\perp \sim Q \lambda$ travel in the $n$ direction, and are described by $n$-collinear fields $\xi^+_{n,p^-,\mathbf{p}_\perp}(x)$ in SCET. The large components of momentum, $p^-$ and $\mathbf{p}_\perp$ are explicit labels on the field, and momentum of order $Q\lambda^2$ is the Fourier transform of the coordinate $x$. This is analogous to use of label-momentum for non-relativistic quarks in NRQCD~\cite{LMR}. Similarly, particles with momenta $p^+ \sim Q$, $p^- \sim Q \lambda^2$ and $\mathbf{p}_\perp \sim Q \lambda$ travel in the $\bn$ direction, and are described by $\bn$-collinear fields $\xi^+_{\bn, p^+, \mathbf{p}_\perp}$. Particles with momenta of order $Q \lambda^2$ are described by ultrasoft fields. In the Breit frame, the outgoing quark is described by a $n$-collinear field, and the incoming quark is described by a $\bn$-collinear field. The choice of coordinate axes is such that the $\perp$ components of label momentum are zero. The fields will be referred to as $\xi_{n}$, $\xi_{\bar n}$ for simplicity.

The Breit frame is the natural frame to use to describe deep inelastic scattering near $x=1$. The power counting automatically implies that $1-x \to 0$, by Eq.~(\ref{2.10}). Nevertheless, it is instructive to also give results in the target rest frame. The target frame is the best frame to compare deep inelastic scattering with $B \to X_s \gamma$. The target rest frame is also the natural frame to use for generic values of $x$, and is a $x$-independent frame. The boost to the Breit frame depends on $x$, though for $x \approx 1$, the boost factor is approximately constant. In the target rest frame, the incoming quark has momentum of order $\lqcd$, and is described by an ultrasoft field. The outgoing quark has momentum components $p_X^+ \sim \lambda \lqcd/Q$, $p_X^- \sim Q^2/\lqcd$, $p_X^\perp \sim Q \lambda$. The outgoing quark turns into a jet moving in the $n$ direction, and so the outgoing particle can be described by a $n$-collinear field.

\section{Matching the current at $Q^2$ from QCD to SCET \label{sec:matchQ}}

The electromagnetic current in QCD is matched onto the SCET current at the scale $Q$. The QCD current is the operator $\bar \psi \gamma^\mu \psi$, and the SCET current is $\bar \xi_n W_n \gamma^\mu W_{\bn}^\dagger \xi_{\bn}$ in the Breit frame, and $\bar \xi_n W_n \gamma^\mu \psi $ in the target rest frame. Here $W_{n,\bn}$ are collinear Wilson lines which are required by collinear gauge invariance. The matching condition will be computed in (a) pure dimensional regularization, i.e.\ using dimensional regularization to regulate both the ultraviolet and infrared divergences, and (b)  by using dimensional regularization for the ultraviolet divergences and off-shellness for the infrared divergences.

\begin{figure}
\includegraphics[width=3cm]{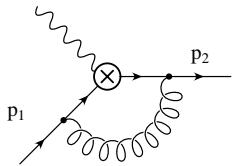}
\caption{One-loop vertex correction to the electromagnetic current in QCD. \label{fig:1}}
\end{figure}

The one-loop vertex graph for the electromagnetic current in QCD is shown in Fig.~\ref{fig:1}, where $p_1$ is the incoming quark momentum, and $p_2=p_1+q$ is the outgoing quark momentum. The QCD one-loop graph
gives
\begin{eqnarray}
V &=&-i g^2C_F  \mu^{2\epsilon} \int { {\rm d}^d k \over (2 \pi )^d} 
\gamma^\alpha { \xslash {p}_2 -\xslash k \over (k - p_2 )^2} \gamma^\mu
{ \xslash{p}_1  - \xslash{k} \over (k- p_1 )^2}
\gamma_\alpha {1 \over k^2}\ ,\nn
\label{4.01}
\end{eqnarray} 
where $C_F=4/3$ is the Casimir of the fundamental representation.

We first consider the computation in pure dimensional regularization, which greatly simplifies the computation of matching conditions in effective field theories. In pure dimensional regularization, the matching coefficient is obtained by computing the finite parts of on-shell diagrams and dropping all the $1/\epsilon$ terms, regardless of whether they arise from ultraviolet or infrared divergences~\cite{AM}. The reason this procedure works is that the ultraviolet divergences in the full and effective theories are canceled by the counterterms in the respective theories. The remaining $1/\epsilon$ terms are infrared divergences, which must agree between the full and effective theory. The $1/\epsilon$ terms cancel in the matching condition, which is the difference between the full and effective theory. Thus the matching condition is the difference of the finite parts of the full and effective theory computation. There is one additional simplification --- on-shell graphs in the effective theory are usually
scaleless integrals, which vanish in pure dimensional regularization, and so have no finite part~\cite{AM}. This eliminates the need to compute the effective theory graphs to determine the matching condition, which is given by the finite part of the full theory graphs.

We will compute Eq.~(\ref{4.01}) keeping the $1/\epsilon$ terms to compare with the matching computation using an infrared regulator. The incoming and outgoing quarks in Fig.~\ref{fig:1} have invariant masses that vanish in the limit $\lambda \to 0$, so the matching coefficient is obtained by evaluating the graph on-shell with $p_1^2=p_2^2=0$. Evaluating the integral in $d=4-2\epsilon$ dimensions gives
\begin{eqnarray}
V &=&  {\alpha_s \over 4 \pi}C_F \gamma^\mu \Biggl[ {1 \over \euv}
-{2 \over \eir^2} - {2 \ln{\mu^2 \over Q^2}+ 4  \over \eir } \nn
&&-\ln^2{\mu^2 \over Q^2} -3 \ln{\mu^2 \over Q^2}  -8 +{\pi^2 \over 6} \Biggr]  ,
\label{4.02}
\end{eqnarray}
where we have distinguished the infrared and ultraviolet divergences by the subscript on $\epsilon$.  However, it is important to keep in mind that all $\epsilon$'s are equal. The integral has a $1/\eir^2$ infrared divergence arising from a combination of soft and collinear divergences. It is this double divergence that leads to the Sudakov double-logarithmic behavior in the endpoint region. In pure dimensional regularization, the wavefunction graphs are scaleless,
\begin{eqnarray}
I_w &=& {\alpha_s \over 4 \pi}C_F \, i \xslash{p}\left[ {1 \over \epsilon_{\text{UV}}} 
-{1 \over \epsilon_{\text{IR}}} \right] ,
\label{4.03}
\end{eqnarray}
and vanish. The net on-shell matrix element of the electromagnetic current in the full theory is the difference of Eqs.~(\ref{4.02},\ref{4.03}) plus the counterterms, which gives (including the tree-graph)
\begin{eqnarray}
\me{p_2}{j^\mu}{p_1} &=& \gamma^\mu \Biggl[ 1 + {\alpha_s \over 4 \pi} C_F \Biggl(
-{2 \over \eir^2} - {2 \ln{\mu^2 \over Q^2}+ 3  \over \eir} \nn
&& -\ln^2{\mu^2 \over Q^2} -3 \ln{\mu^2 \over Q^2}  -8 +{\pi^2 \over 6} \Biggr) \Biggr] + \text{c.t.}\ . \nn
\label{4.04}
\end{eqnarray}
The $1/\euv$ terms cancel, so there is no counterterm. This cancellation is required, since the electromagnetic current is a conserved current and has no anomalous dimension in QCD. The graphs in the effective theory are all scaleless, and vanish in dimensional regularization. The matching coefficient of the current in the effective theory is the finite part of Eq.~(\ref{4.04}),
\begin{eqnarray}
C(\mu) &=&1 +  {\alpha_s (\mu) \over 4 \pi} C_F \left[ 
-\ln^2{\mu^2 \over Q^2} -3 \ln{\mu^2 \over Q^2}  -8 +{\pi^2 \over 6} \right] .\nn
\label{4.06}
\end{eqnarray}
The $1/\eir$ terms in Eq.~(\ref{4.04}), which are the negative of the $1/\euv$ terms in the effective theory, give the anomalous dimension of the current in the effective theory, as we will see in the next section.

The logarithms in the matching coefficient $C(\mu)$ can be minimized by choosing the matching scale $\mu=Q$, at which
\begin{eqnarray}
C(Q) &=&1 + {\alpha_s(Q) \over 4 \pi}C_F  \left[   -8 +{\pi^2 \over 6} \right] .
\label{4.10}
\end{eqnarray}

The matching computation can be repeated by regulating the infrared divergence by using off-shell initial and final states, with $p_1^2=p_2^2\not=0$. The graph in Fig.~\ref{fig:1} gives
\begin{eqnarray}
V&=& {\alpha_s \over 4 \pi} C_F\gamma^\mu
\Biggl[{1\over \euv} - \ln {Q^2 \over \mu^2} -2 \ln{p_1^2 \over Q^2} \ln{p_2^2 \over Q^2}\nn
&&-2 \ln{p_1^2 \over Q^2}-2 \ln{p_2^2 \over Q^2}-{2 \pi^2\over 3} \Biggr] ,
\label{4.07}
\end{eqnarray}
where the $1/\epsilon$ term is purely an ultraviolet divergence, since the infrared divergences have been regulated by the off-shellness. The evaluation of Eq.~(\ref{4.07}) is considerably more complicated than that of Eq.~(\ref{4.02}). The wavefunction graph is
\begin{eqnarray}
I_w&=& {\alpha_s \over 4 \pi} C_F \, i \xslash{p}\left[{1\over \epsilon_{\text{UV}}} + 1 - \ln{-p^2 \over  \mu^2 } \right] ,
\label{4.08}
\end{eqnarray}
so that the matrix element in the full theory including  the tree-graph is
\begin{eqnarray}
\me{p_2}{j^\mu}{p_1} &= &\gamma^\mu\Biggl[ 1+C_F {\alpha_s \over 4 \pi} 
\Biggl( - \ln {Q^2 \over \mu^2} -2 \ln{p_1^2 \over Q^2} \ln{p_2^2 \over Q^2}\nn
&&-2 \ln{p_1^2 \over Q^2}-2 \ln{p_2^2 \over Q^2}  +\frac 1 2 \ln{-p_1^2 \over  \mu^2 } \nn
&& + \frac 1 2 \ln{-p_2^2 \over  \mu^2 }  -1-{2 \pi^2\over 3} \Biggr)\Biggr]
+\text{c.t.}\ .
\label{4.09}
\end{eqnarray}
The ultraviolet counterterm vanishes as before, as it must, since it does not depend on the choice of infrared regulator. The matching condition is given by subtracting from Eq.~(\ref{4.09}) the matrix element in the effective theory. The effective theory integrals are no longer scaleless, since they depend on $p_i^2$, and must be evaluated to obtain the matching condition if an off-shellness is used to regulated the infrared divergence. The matrix element in the effective theory with an off-shellness is given in the next section, where the anomalous dimension of the SCET current is computed. Taking the difference of the result, Eq.~(\ref{5.11}), and Eq.~(\ref{4.09}) gives the same matching condition as before, Eq.~(\ref{4.06}). The computation of Eq.~(\ref{4.06}) is clearly simpler using dimensional regularization to regulate the infrared divergence, since it does not require the effective theory computation at this stage, leaving it to the next section where it properly belongs.

\section{Anomalous dimension of the SCET current \label{sec:Qrun}}

The electromagnetic current in QCD is matched at the scale $Q$ onto the SCET current. In the Breit frame, the current is
\begin{eqnarray}
j^\mu &=& C(Q)\ \bar \xi_{n,Q} W_n \gamma^\mu W^\dagger_{\bn} \xi_{\bn, Q} ,
\label{5.02}
\end{eqnarray}
where the $n$-collinear quark has label momentum $\bn \cdot p = Q$, $\mathbf{p}_\perp=0$, and the $\bar n$-collinear quark has label momentum $n \cdot p = Q$, $\mathbf{p}_\perp=0$. $C(Q)$ is the matching coefficient computed in Eq.~(\ref{4.10}). 

In the target rest frame, the current is
\begin{eqnarray}
j^\mu &=& C(Q)\ \bar \xi_{n,q^-} W_n \gamma^\mu \psi_u ,
\label{5.01}
\end{eqnarray}
where $\psi_u$ is an ultrasoft quark, $W_n$ is a collinear Wilson line, and $ \xi_{n,q^-}$ is a $n$-collinear quark field with labels $\bn \cdot p =q^-$, $\mathbf{p}_\perp=0$. 

\subsection{Breit Frame}

\begin{figure*}
\def\size{4.5 cm}
\hbox{\hspace{2.25cm}\vbox{\hbox to \size {\hfil \includegraphics[width=3cm]{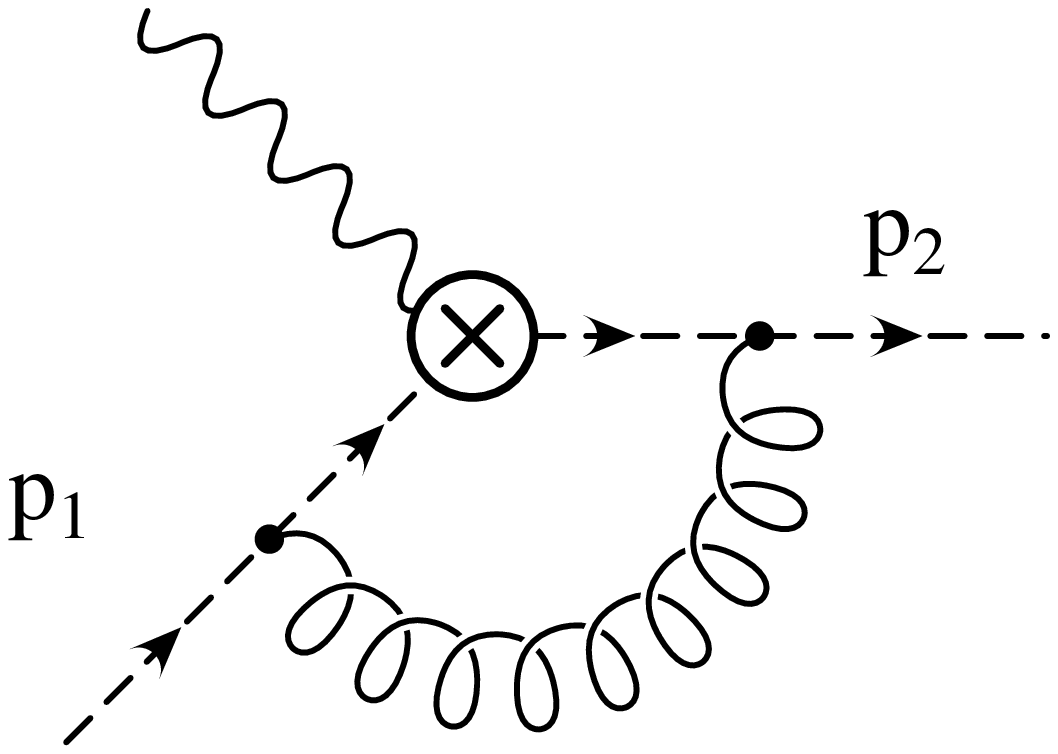} \hfil }\hbox to \size {\hfil(a)\hfil}}
\vbox{\hbox to \size {\hfil \includegraphics[width=3cm]{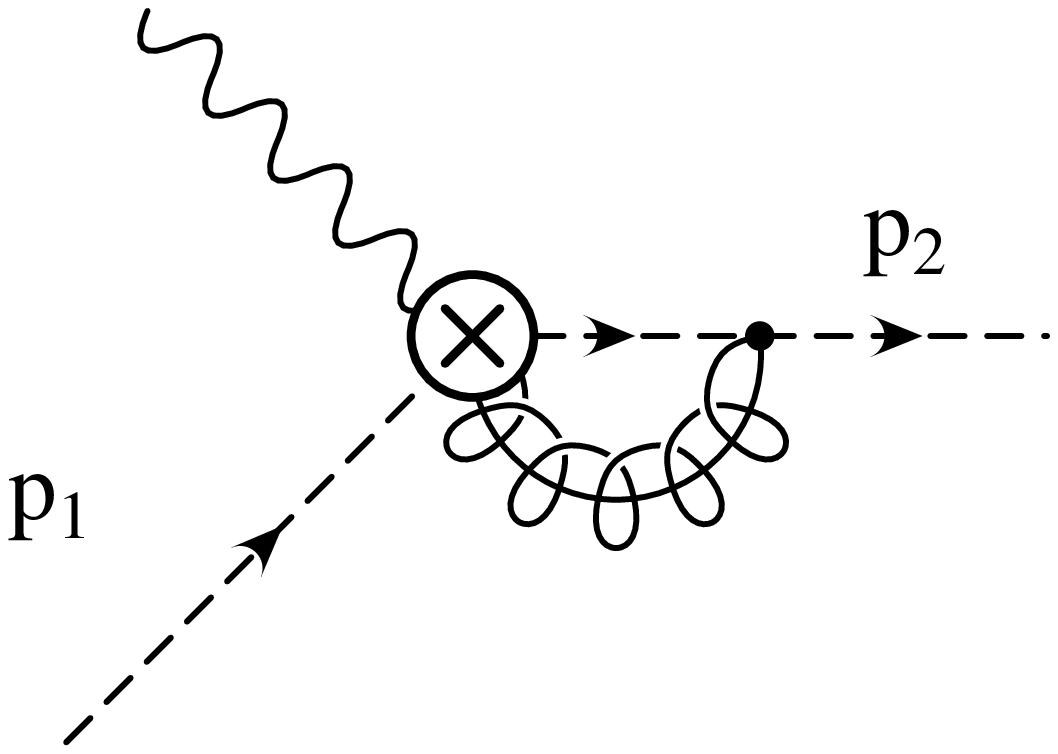} \hfil}\hbox to \size {\hfil(b)\hfil}}
\vbox{\hbox to \size {\hfil \includegraphics[width=3cm]{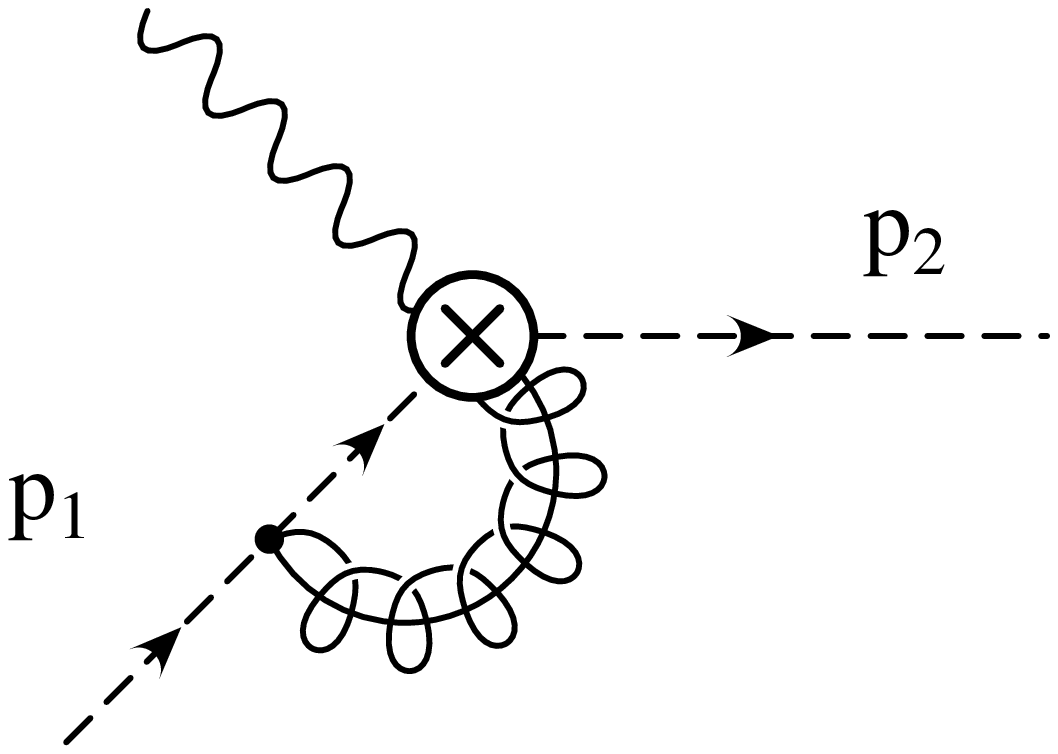} \hfil}\hbox to \size {\hfil(c)\hfil}}}
\caption{One loop correction to the electromagnetic vertex in the Breit frame from
(a) ultrasoft, (b) $n$-collinear and (c) $\bn$-collinear gluons. \label{fig:2}}
\end{figure*}

The one-loop anomalous dimension of the SCET current in the Breit frame is given by the graphs in Figs.~\ref{fig:2}, as well as wavefunction renormalization graphs.

The ultrasoft graph, Fig.~\ref{fig:2}(a), gives
\begin{eqnarray}
I_s&=&-i g^2 C_F \mu^{2\epsilon} \int { {\rm d}^d k \over (2 \pi )^d} 
\ n^\alpha {1  \over n \cdot (p_2-k)+i 0^+} \gamma^\mu \nn
&& \times { 1 \over \bar n \cdot (p_1 -k )+i0^+}
\bar n_\alpha {1 \over k^2+i0^+}\ .
\label{5.03}
\end{eqnarray} 
Doing the $k^+$ integral by contours and using the substitution $k^- = x p_1^-$ gives
\begin{eqnarray}
I_s &=& -{g^2} C_F \mu^{2\epsilon} 
\int_{0}^\infty {dx \over 2 \pi} \int { {\rm d}^{d-2} k_\perp \over (2 \pi )^{d-2}} \nn
&&\times  { \gamma^\mu \over
\left[ 1-x+ i 0^+ \right]
\left[p_2^+ p_1^- x - \mathbf{k}_\perp^2 + i0^+ \right]}\nn
&=&{g^2 \over 8 \pi^2}C_F \gamma^\mu  \mu^{2\epsilon} 
\int_{0}^\infty {dx} 
 { \Gamma(\epsilon)\left[-p_2^+ p_1^- x\right]^{-\epsilon} \over
\left[ 1-x+ i 0^+ \right]}  \nn
&=&-{g^2 \over 8 \pi^2}C_F \gamma^\mu \mu^{2\epsilon}  
 \Gamma(\epsilon)\left[p_2^+ p_1^- \right]^{-\epsilon}  \pi
 \csc \epsilon \pi \nn
&=&-{g^2 \over 8 \pi^2} C_F\left[{1\over \epsilon^2} - {1\over \epsilon} \ln{p_2^+ p_1^- \over \mu^2} + \frac 12 \ln^2{p_2^+ p_1^- \over \mu^2}  + {\pi^2 \over 4}
\right] .\nn
\label{5.04}
\end{eqnarray} 

The $n$-collinear gluon graph Fig.~\ref{fig:2}(b) gives
\begin{eqnarray}
I_n&=&-i g^2 C_F  \mu^{2\epsilon} \int { {\rm d}^d k \over (2 \pi )^d} 
{\xslash{\bar n}\ n^\alpha \over 2}  {\xslash{n}\ \bar n \cdot (p_2-k)  \over 2 (p_2 - k )^2+i0^+} \gamma^\mu\nn
&& \times { 1 \over  - \bar n \cdot k+i0^+} \bar n_\alpha {1 \over k^2+i0^+} \nn
&=& -2 i g^2  C_F \mu^{2\epsilon} \int { {\rm d}^d k \over (2 \pi )^d} 
  {\bar n \cdot (p_2-k)  \over (p_2 - k )^2+i0^+} \gamma^\mu\nn
&& \times { 1 \over - \bar n \cdot k+i0^+}  {1 \over k^2+i0^+} \ .
\label{5.05}
\end{eqnarray} 
Evaluating the $k^+$ integral by contours, and letting $k^-=z p_2^-$ gives
\begin{eqnarray}
I_{n} &=&  g^2  C_F \mu^{2\epsilon} \gamma^\mu
\int_0^{1} {dz  \over 2 \pi} \int { {\rm d}^{d-2} k_\perp \over (2 \pi )^{d-2}}  {  (1-z)  
  \over  z  \left[ z (1-z) p_2^2  -  \mathbf{k}_\perp^2 \right] } \nonumber\\[5pt]
&=&  -{g^2 \over 8 \pi^2}   \mu^{2\epsilon} \gamma^\mu
\int_0^{1} {dz } \Gamma(\epsilon) {  (1-z)  \left[ -p_2^2 z(1-z)\right]^{-\epsilon}
  \over  z } \nn
&=& - {g^2 \over 8 \pi^2}C_F   \mu^{2\epsilon} \gamma^\mu
{ \Gamma(\epsilon) \Gamma(-\epsilon) \Gamma(2-\epsilon)
  \over  \Gamma(2-2\epsilon) }\left[ -p_2^2\right]^{-\epsilon} \nn
&=&  -{g^2 \over 8 \pi^2}  C_F \gamma^\mu
\Bigl[-{1\over \epsilon^2} - {1 \over \epsilon} + {1 \over \epsilon} \ln {-p_2^2 \over \mu^2} \nn
&&\qquad - \frac 12  \ln^2 {-p_2^2 \over \mu^2} +\ln {-p_2^2 \over \mu^2}
-2 + {\pi^2 \over 12 }\Bigr] .
\label{5.06}
\end{eqnarray} 

The $\bar n$-collinear gluon graph  Fig.~\ref{fig:2}(c) gives
\begin{eqnarray}
I_{\bn} &= &-i g^2C_F   \mu^{2\epsilon} \int { {\rm d}^d k \over (2 \pi )^d} 
n^\alpha { 1 \over  - n \cdot k+i0^+} \gamma^\mu \nn
&& \times  {\xslash{\bar n}\  n \cdot (p_1-k)  \over 2 (p_1 - k )^2+i0^+}
{\xslash{n}\ \bar n_\alpha \over 2}   
 {1 \over k^2+i0^+} \ ,
\label{5.07}
\end{eqnarray} 
which is Eq.~(\ref{5.06}) with $p_2^2 \to p_1^2$,
\begin{eqnarray}
I_{\bn}
&=&  -{g^2 \over 8 \pi^2}C_F   \gamma^\mu
\Bigl[-{1\over \epsilon^2} - {1 \over \epsilon} + {1 \over \epsilon} \ln {-p_1^2 \over \mu^2} \nn
&&\qquad - \frac 12  \ln^2 {-p_1^2 \over \mu^2} +\ln {-p_1^2 \over \mu^2}
-2 + {\pi^2 \over 12 }\Bigr] .
\label{5.08}
\end{eqnarray} 

The remaining graphs  are the wavefunction graphs. The ultrasoft gluon contribution to wavefunction renormalization vanishes in Feynman gauge, since $n^2=\bn^2=0$. The collinear wavefunction renormalization graph is the same as the massless quark wavefunction renormalization in QCD, Eq.~(\ref{4.08}), since the interaction of $n$-collinear quarks with $n$-collinear gluons is the same as the interaction of quarks with gluons in full QCD.

The matrix element of the current in the effective theory is given by the sum of Eqs.~(\ref{5.04},\ref{5.06},\ref{5.08}) and subtracting half the wavefunction renormalization Eq.~(\ref{4.08}) for each external quark. The net result is
\begin{eqnarray}
&&\me{p_2}{j^\mu}{p_1}_{\text{bare}}\nn
&=&{\alpha_s \over 4 \pi}C_F \Bigl[{2\over \epsilon^2} + {3 \over \epsilon} - {2 \over \epsilon} \ln{Q^2\over \mu^2}  -\ln^2{Q^2\over \mu^2} \nn
&&+2 \ln{-p_1^2 \over \mu^2} \ln{Q^2\over \mu^2}  + 2\ln{-p_2^2 \over \mu^2} \ln{Q^2\over \mu^2} - 2\ln{-p_1^2 \over \mu^2}\ln{-p_2^2 \over \mu^2} \nn
&&  -\frac 3 2\ln{-p_1^2 \over \mu^2} -\frac 3 2\ln{-p_2^2 \over \mu^2} 
+7  - {5\pi^2 \over 6 }\Bigr] .
\label{5.09}
\end{eqnarray}
The ultraviolet divergence of this result agree with the ultraviolet divergence in the effective theory inferred from Eq.~(\ref{4.04}). Exactly on-shell, the effective theory integrals are scaleless and vanish because the $1/\euv$ ultraviolet divergences cancel the $1/\eir$ infrared divergences. The ultraviolet divergence in the effective theory should therefore be the negative of the $1/\eir$ terms in Eq.~(\ref{4.04}), which agrees with the divergence in Eq.~(\ref{5.09}). The SCET current operator is multiplicatively renormalized, and there is no operator mixing. This allows one to compute the anomalous dimension of the SCET current directly from the $1/\eir$ terms in the full theory matrix element. In cases with operator mixing, the $1/\eir$ terms in the full theory matrix element gives the value of the anomalous dimension matrix times the operators coefficients evaluated at the matching scale.

The infinite part of the matrix element is canceled by the vertex and wavefunction counterterms in the effective theory so that the renormalized matrix element is
\begin{eqnarray}
&&\me{p_2}{j^\mu}{p_1}_{\text{ren}} \nn
&=&{\alpha_s \over 4 \pi}C_F \Bigl[  -\ln^2{Q^2\over \mu^2} +2 \ln{-p_1^2 \over \mu^2} \ln{Q^2\over \mu^2}  + 2\ln{-p_2^2 \over \mu^2} \ln{Q^2\over \mu^2} \nn
&& - 2\ln{-p_1^2 \over \mu^2}\ln{-p_2^2 \over \mu^2}   -\frac 3 2\ln{-p_1^2 \over \mu^2} -\frac 3 2\ln{-p_2^2 \over \mu^2} +7  - {5\pi^2 \over 6 }\Bigr]. \nn
\label{5.11}
\end{eqnarray}
The infrared divergence as $p_i^2 \to 0$ in the full theory calculation, Eq.~(\ref{4.09}) agrees with the infrared divergence of the effective theory calculation, Eq.~(\ref{5.11}), and the difference gives the matching condition, Eq.~(\ref{4.06}), which is free of infrared divergences, and depends only on $Q^2$, not on $p_i^2$.

The ultraviolet counterterm for the ultrasoft graph Fig.~\ref{fig:2}(a), $n$-collinear graph Fig.~\ref{fig:2}(b) and $\bn$-collinear graph Fig.~\ref{fig:2}(c) and wavefunction graph are from Eqs.~(\ref{5.04},\ref{5.06},\ref{5.08},\ref{4.08})
\begin{eqnarray}
\text{ultrasoft:}&&{\alpha_s \over 4 \pi}C_F \left[{2\over \epsilon^2} - {2 \over \epsilon} \ln {p_2^+ p_1^- \over \mu^2}  \right] ,\nn
\text{$n$-collinear:}&&{\alpha_s \over 4 \pi}C_F \left[-{2\over \epsilon^2} - {2 \over \epsilon} +{2\over \epsilon}  \ln {-p_2^2 \over \mu^2}  \right],\nn
\text{$\bn$-collinear:}&&{\alpha_s \over 4 \pi}C_F \left[-{2\over \epsilon^2} - {2 \over \epsilon} +{2 \over \epsilon} \ln {-p_1^2 \over \mu^2}  \right], \nn
\text{wavefunction:}&&{\alpha_s \over 4 \pi}C_F \left[{1 \over \epsilon} \right] ,
\label{5.13}
\end{eqnarray}
respectively. The individual counterterms are sensitive to the infrared through their dependence on the small components of momentum, $p_1^-$ and $p_2^+$. The total counterterm is the sum of the four terms above,
\begin{eqnarray}
\text{c.t.} &=&{\alpha_s \over 4 \pi}C_F \left[-{2\over \epsilon^2}  - {3 \over \epsilon} - {2 \over \epsilon} \ln {p_2^+ p_1^-  \mu^2 \over p_1^2 p_2^2}  \right] ,\nn
&=&{\alpha_s \over 4 \pi}C_F \left[-{2\over \epsilon^2}  - {3 \over \epsilon} - {2 \over \epsilon} \ln {  \mu^2 \over p_2^- p_1^+}  \right] ,\nn
&=&{\alpha_s \over 4 \pi}C_F \left[-{2\over \epsilon^2}  - {3 \over \epsilon} - {2 \over \epsilon} \ln {  \mu^2 \over Q^2}  \right] ,
\label{5.14}
\end{eqnarray}
and depends only on the label momenta $p_2^-$ and $p_1^+$ which are both $Q$.
The counterterms Eq.~(\ref{5.14}) give the anomalous dimension for the coefficient of the current in the effective theory,
\begin{eqnarray}
 \mu {{\rm d} C(\mu) \over {\rm d} \mu}  &=& \gamma_1(\mu)\ C(\mu)\ ,\nn
\gamma_1(\mu)   &=& - C_F {\alpha_s(\mu) \over 4 \pi} \left[ 4 \ln {\mu^2 \over Q^2} + 6 \right]\ .
\label{5.12}
\end{eqnarray}
The SCET current anomalous dimension depends on $\ln \mu$, because the one-loop diagrams have $1/\epsilon^2$ terms from combined collinear and soft divergences. Consistency of the effective theory implies that to all orders, the anomalous dimension is at most linear in $\ln \mu$, as will be shown in Sec.~\ref{sec:gamma}.

The cancellation of $1/\epsilon \ln p_1^-$ and $1/\epsilon \ln p_2^+$ between the soft and collinear graphs might suggest that the coefficients in the two sectors must be the same, i.e.\ that the two contributions must have the same value of $\alpha_s$. This suggests that SCET should use two-stage running, in which all coupling constants are evaluated at a single scale $\mu$, unlike NRQCD, which requires one-stage running using the velocity renormalization group~\cite{LMR,IS}. In NRQCD, there is a cancellation of infrared divergences between the soft and ultrasoft sectors that naively suggests that both should have the same value of $\alpha_s$. However, this is not the case, and a proper treatment of NRQCD has the soft coupling constant evaluated at the scale $m\nu$ and ultrasoft coupling constant evaluated at the scale $m \nu^2$~\cite{LMR,IS}. It has been pointed out that one-stage and two-stage running give the same result in SCET for quantities which have been computed so far~\cite{Fleming}. In NRQCD, the difference between one- and two-stage running first occurs at order $v^2$ in the power counting. SCET anomalous dimensions have so far been computed only to leading order in $\lambda$, and do not distinguish between one-stage and two-stage running.

\subsection{Target Rest Frame}

\begin{figure}
\def\size{4.5 cm}
\hbox{\vbox{\hbox to \size {\hfil \includegraphics[width=3cm]{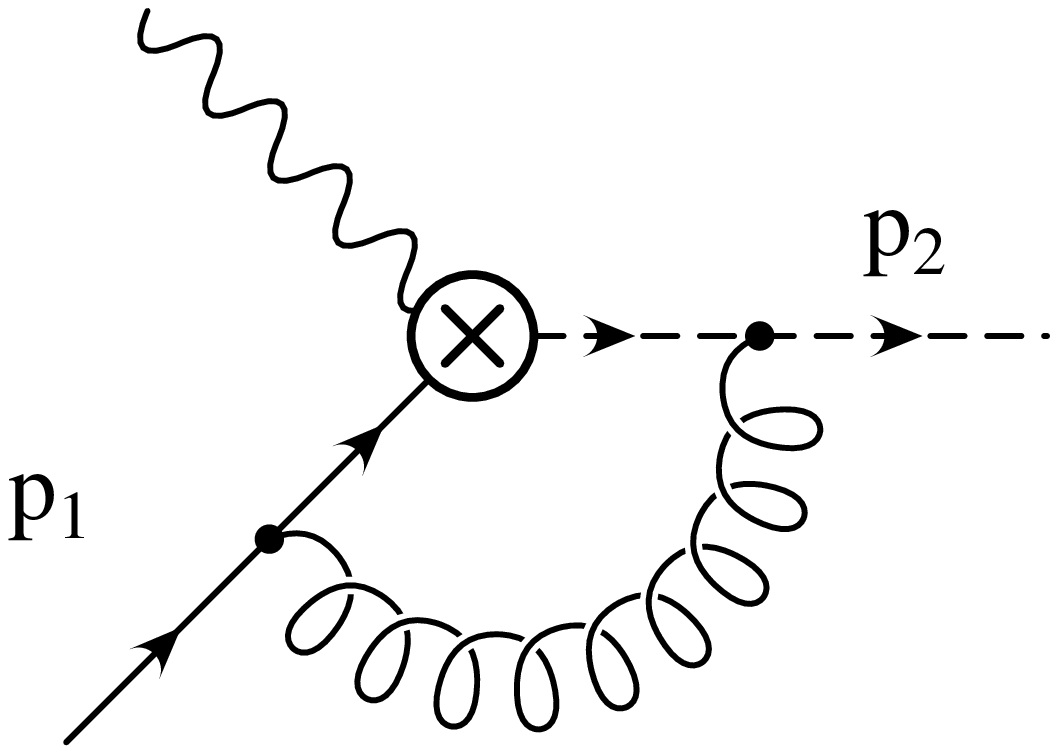} \hfil }\hbox to \size {\hfil(a)\hfil}}
\vbox{\hbox to \size {\hfil \includegraphics[width=3cm]{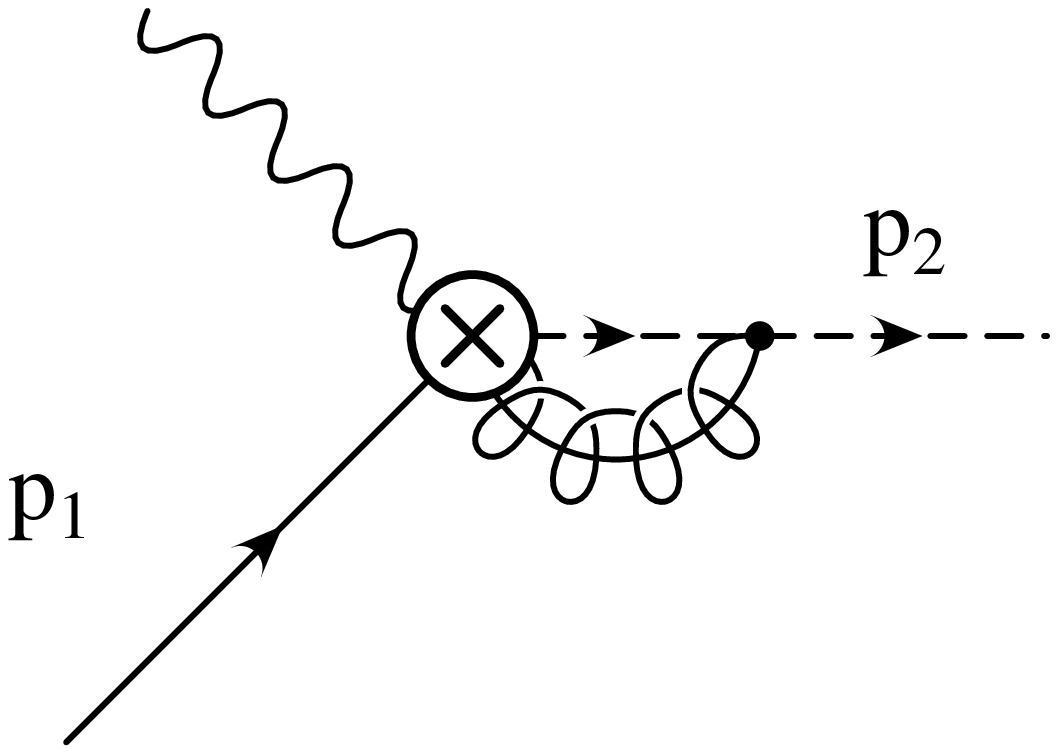} \hfil}\hbox to \size {\hfil(b)\hfil}}
}
\caption{One loop correction to the electromagnetic vertex in the target rest frame from
(a) ultrasoft and (b) $n$-collinear gluons. \label{fig:3}}
\end{figure}

In the target rest frame, the electromagnetic current in the effective theory contains a $n$-collinear quark and an ultrasoft quark. The diagrams in the effective theory in the target rest frame are those in Fig.~\ref{fig:3} and wavefunction graphs.
Note that there is no $\bn$-collinear graph, since the current Eq.~(\ref{5.02}) does not contain $W_{\bn}$ in the target rest frame.

The $n$-collinear graph Fig.~\ref{fig:3}(b) is the same as in the Breit frame, and is given by Eq.~(\ref{5.06}). The result is frame-independent, since the final expression only depends on $p_2^2$, not on the individual components $p_2^\pm$ or the momentum $p_1$ of the incoming quark. The non-zero wavefunction renormalization graphs also agree between the two theories. The $n$-collinear wavefunction renormalization is the same as the corresponding graph in the Breit frame, and the ultrasoft wavefunction renormalization of the ultrasoft quark is the same as the $\bn$-collinear wavefunction renormalization of the $\bn$ quark, since both are equal to wavefunction renormalization of a massless quark in QCD.  Each wavefunction graphs depends only on $p^2$ for a single particle.

The remaining graph is the ultrasoft graph Fig.~\ref{fig:3}(a). The ultrasoft graph in the target rest frame must contain the contributions of both the ultrasoft and $\bn$-collinear graphs in the Breit frame. The ultrasoft graph in the target rest frame is
\begin{eqnarray}
I&=&-ig^2 C_F \mu^{2\epsilon} \int { {\rm d}^d k \over (2 \pi )^d} 
n^\alpha {1  \over n \cdot (p_2-k)+i0^+} \gamma^\mu \nn
&& \times {  \xslash{p}_1 - \xslash k \over (p_1 -k )^2+i0^+}
\gamma_\alpha {1 \over k^2+i0^+}\nn
&=&-ig^2 C_F \mu^{2\epsilon} \nn
&&\int { {\rm d}^d k \over (2 \pi )^d} 
 { \gamma^\mu  (\xslash{p}_1 - \xslash k ) \xslash{n} \over [n \cdot (p_2-k)+i0^+]
(p_1 -k )^2+i0^+} {1 \over k^2+i0^+}\ . \nn
\label{5.15}
\end{eqnarray} 
Doing the $k^+$ integral, and using $k^+=zp_1^+$ gives
\begin{eqnarray}
I&=&-{g^2}C_F \mu^{2\epsilon} \gamma^\mu
\int_0^{1} {dz  \over 2 \pi}
 \int { {\rm d}^{d-2} k_\perp \over (2 \pi )^{d-2}} \nn
&&
\times {  p_1^+(1-z) \over (p_2^+-p_1^+ z)
(z(1-z)p_1^2 -\mathbf{k}_\perp^2)} \nn
&=& {g^2 \over 8\pi^2}C_F \mu^{2\epsilon} \gamma^\mu \Gamma(\epsilon)
\int_0^{1} {dz} {  p_1^+(1-z) \over (p_2^+-p_1^+ z)}\left[-p_1^2 z(1-z)\right]^{-\epsilon}\ .\nn
\label{5.16}
\end{eqnarray}
The ratio $p_2^+/p_1^+$ is
\begin{eqnarray}
{p_2^+ \over p_1^+} = 1-x = {p_2^2 \over p_1^+ p_2^-} = {x p_2^2\over Q^2} \approx {p_2^2 \over Q^2}
=\mathcal{O}\left( \lambda \right)\ .
\label{5.18}
\end{eqnarray}
In SCET, terms of order $1-x$ are of order the expansion parameter $\lambda$ in the power counting. Equation~(\ref{5.16}) can be evaluated in the limit $p_2^+/p_1^+ \to 0$, which simplifies the computation, and gives
\begin{eqnarray}
I &=& {g^2 \over 8\pi^2} C_F \gamma^\mu
\Biggl\{  \left[{1\over \epsilon} -\ln{-p_1^2 \over \mu^2}\right]
 \left(1 + \ln {-p_2^+ \over p_1^+ } \right)  \nn
&&+ \left(2 - {\pi^2 \over 3} - \frac 12 \ln^2 {-p_2^+ \over p_1^+ }  \right)\Biggr\} .
\label{5.17}
\end{eqnarray} 
This result is identical to the sum of the ultrasoft and $\bn$-collinear graphs in the Breit frame, given in Eqs.~(\ref{5.04},\ref{5.08}).\footnote{The components of $p_i$ have different values in the two frames, but $p_i^2$ and $p_2^+/p_1^+$ are equal, since the Breit frame is obtained from the target frame by a boost along the $z$ axis.} The ultrasoft graph in the target rest frame has no $1/\epsilon^2$ divergence. Since it is the sum of the  ultrasoft and $\bn$-collinear graphs in the Breit frame, these graphs must have $1/\epsilon^2$ terms of opposite sign, which agrees with the explicit one-loop computation in Eqs.~(\ref{5.04},\ref{5.08}). This cancellation is expected to persist at higher orders.

 The $1/\epsilon$ term in Eq.~(\ref{5.17}) depends on the infrared regulator through $p_2^2$, using Eq.~(\ref{5.18}). This infrared dependence is canceled by the $n$-collinear graph, which can only depend on $p_2^2$. The analog of Eq.~(\ref{5.13}) is
\begin{eqnarray}
\text{ultrasoft:}&&{\alpha_s \over 4 \pi}C_F \left[-{2\over \epsilon} - {2\over \epsilon} \ln {-p_2^+ \over p_1^+}  \right] ,\nn
\text{$n$-collinear:}&&{\alpha_s \over 4 \pi}C_F \left[-{2\over \epsilon^2} - {2 \over \epsilon} +{2\over \epsilon}  \ln {-p_2^2 \over \mu^2}  \right],\nn
\text{wavefunction:}&&{\alpha_s \over 4 \pi}C_F \left[{1 \over \epsilon} \right] ,
\label{5.13a}
\end{eqnarray}
so that the total counterterm is
\begin{eqnarray}
\text{c.t.} &=&{\alpha_s \over 4 \pi}C_F \left[-{2\over \epsilon^2}-{3\over \epsilon} - {2\over \epsilon} \ln {p_2^+ \mu^2 \over p_1^+ p_2^2}  \right],\nn
&=&{\alpha_s \over 4 \pi}C_F \left[-{2\over \epsilon^2}-{3\over \epsilon} - {2\over \epsilon} \ln { \mu^2 \over p_1^+ p_2^-}  \right],
\label{5.14a}
\end{eqnarray}
which is the same as Eq.~(\ref{5.14}), and leads to the same anomalous dimension,
Eq.~(\ref{5.12}).

The on-shell matrix element in the target rest frame has the same value as in the Breit frame, and leads to the same matching condition Eq.~(\ref{4.06}).

\section{Matching at $Q^2(1-x)$ onto the parton distribution function \label{sec:matchQl}}

At the scale $Q^2(1-x)$, the invariant mass of the final hadronic state $p_X^2$ can be treated as large, and the final state can be integrated out. This is done by integrating out the $n$-collinear modes from the effective theory.  Since the current Eq.~(\ref{5.02},\ref{5.01}) contains $\xi_n$ fields which are integrated out, one matches the product of two currents, rather than a single current, onto the effective theory below $Q^2(1-x)$.  The matching coefficients are determined by computing the matrix elements of
\begin{eqnarray}
W^{\mu \nu} = {1\over 2 \pi} \int {\rm d}^4 x e^{i q \cdot x} j^\mu(x) j^\nu(0),
\label{6.01}
\end{eqnarray}
at fixed $x$ and $q^2$. Note that we have a product of currents, rather than a time-ordered product. The matrix element of the product of currents is given by the discontinuity in the matrix element of the time-ordered product.\footnote{The discontinuity of a diagram containing $i0^+$ terms is defined by taking the difference of the diagram, and the diagram with $i0^+ \to i0^-$.} We will use this procedure, since time-ordered products can be computed using conventional Feynman diagram perturbation theory.

\subsection{Target Rest Frame}

\begin{figure}
\includegraphics[width=4cm]{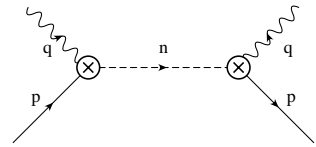}
\caption{Tree graph for the product of two currents in the target rest frame. \label{fig:4a}}
\end{figure}

The tree graph for the product of two currents with $\xi_n$ integrated out is shown in Fig.~\ref{fig:4a}.
The spin averaged matrix element of the tree level graph is
\begin{eqnarray}
&&\text{Disc}\ {1\over 2 \pi} {1 \over 2}\, \text{Tr}\ \xslash{p}\gamma^\nu {i\xslash n \over 2}
{\bar n \cdot (p+q) \over (p+q)^2+i 0^+}  \gamma^\mu \nn
&=&\text{Disc}\ {i\over 2 \pi}T^{\mu \nu} {p^+ (p^-+q^-) \over (p+q)^2 + i 0^+} \nn
&\approx&T^{\mu \nu} \delta\left(1 + q^+/p^+ \right),
\label{6.02}
\end{eqnarray}
where
\begin{eqnarray}
T^{\mu \nu} &=& - g_{\mu \nu} + { p^\mu n^\nu + p^\nu n^\mu \over n \cdot p}.
\label{6.03}
\end{eqnarray}
Since $p$ is at rest, $p^\mu \propto v^\mu=(1,0,0,0)$
\begin{eqnarray}
T^{\mu \nu} &=& - g_{\mu \nu} + { v^\mu n^\nu + v^\nu n^\mu},
\label{6.10}
\end{eqnarray}
independent of the momentum of the target.

Define the quark distribution operator by~\cite{Collins}
\begin{eqnarray}
O_q(r^+) &=& {1 \over 4 \pi} \int_{-\infty}^\infty \diff \xi e^{-i \xi r^+}
 \bar \psi_u \! \left(n \xi \right)\xslash{n} Y\left(n \xi,0 \right)  \psi_u \! \left( 0 \right), \nn
\label{6.04}
\end{eqnarray}
where $Y$ is an eikonal Wilson line from $0$ to $n \xi$ containing ultrasoft gluons $A_u$,
\begin{eqnarray}
Y\left(n \xi,0 \right) &=& P \exp \left[- i g  \int_0^{\xi} n \cdot A_u(n z)\ {\rm d} z \right]
\label{6.04a}
\end{eqnarray}
and $\psi_u$ are ultrasoft quark fields. The Feynman rules are given by taking the discontinuity of the diagram, since the operator is a product, not a time-ordered product. The quark distribution in a target $T$ with momentum $P$ is defined by~\cite{Collins}
\begin{eqnarray}
f_{q/T} (x) &=& \bra{T,P} O_q\left(x P^+ \right) \ket{T,P} \ .
\label{6.05}
\end{eqnarray}
The only difference between the quark distribution operator Eq.~(\ref{6.04}) and the conventional Collins-Soper definition is the replacement of the full theory quark field by the ultrasoft quark field in SCET.

\begin{figure}
\includegraphics[width=4cm]{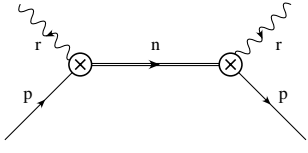}
\caption{Quark matrix element of the quark distribution operator. \label{fig:4z}}
\end{figure}

The spin-averaged tree-level matrix element of the quark distribution operator is (see Fig.~\ref{fig:4z})
\begin{eqnarray}
&&\text{Disc}\ {1\over 4 \pi} {i \over n \cdot \left( p - r \right)  } \frac 12 \text{Tr}\ \xslash{n} \xslash{p} \nn
&&\text{Disc}\ {1\over 4 \pi} {2 i p^+ \over p^+ - r^+ + i 0^+} \nn
&=&  \delta\left( 1-r^+/p^+\right),
\label{6.06}
\end{eqnarray}
so the tree level relation is
\begin{eqnarray}
W^{\mu \nu} &=& T^{\mu \nu} O_q(-q^+)\ .
\label{6.07}
\end{eqnarray}
This is an operator relation independent of the matrix element since $T^{\mu\nu}$ in Eq.~(\ref{6.10}) can be written in a form independent of the target state. The minus sign in the argument of $O_q$ arises because $q$ is an incoming momentum and $r$ is an outgoing momentum. Equation~(\ref{6.07}) can also be written as a convolution
\begin{eqnarray}
W^{\mu \nu} &=& T^{\mu \nu} \int {{\rm d} r^+ \over r^+} \delta\left(1- {-q^+ \over r^+}\right)
O_q(r^+) \ .
\label{6.08}
\end{eqnarray}
It is convenient to write $q^+=-y p^+$, and $r^+=w p^+$
so that  Eq.~(\ref{6.08}) becomes
\begin{eqnarray}
W^{\mu \nu}(q^+=-yp^+) &=& T^{\mu \nu} \int {{\rm d}w \over w} \delta\left(1-{y \over w} \right) O_q(w p^+) \ .\nn
\label{6.08a}
\end{eqnarray}
Note that $y$ and $w$ in this equation are defined with respect to the parton momentum $p$ rather than the hadron momentum $P$.

The analysis has been restricted to spin-independent structure functions for simplicity. It is straightforward to generalize this to spin-dependent structure functions, which involve spin-dependent quark distribution operators~\cite{spin}.

\begin{figure*}
\def\size{5.75 cm}
\hbox{\vbox{\hbox to \size {\hfil \includegraphics[width=4cm]{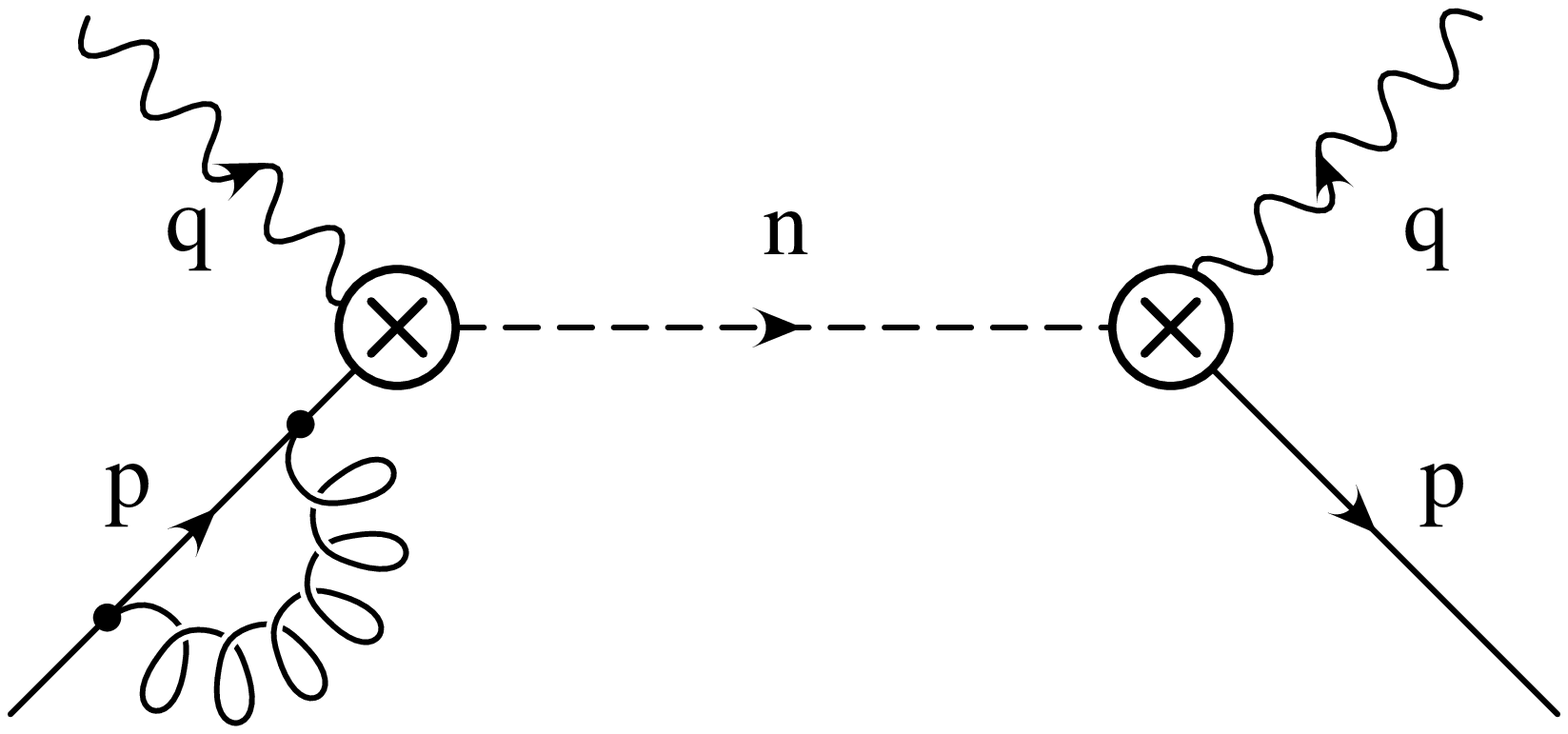} \hfil }\hbox to \size {\hfil(a)\hfil}}
\vbox{\hbox to \size {\hfil \includegraphics[width=4cm]{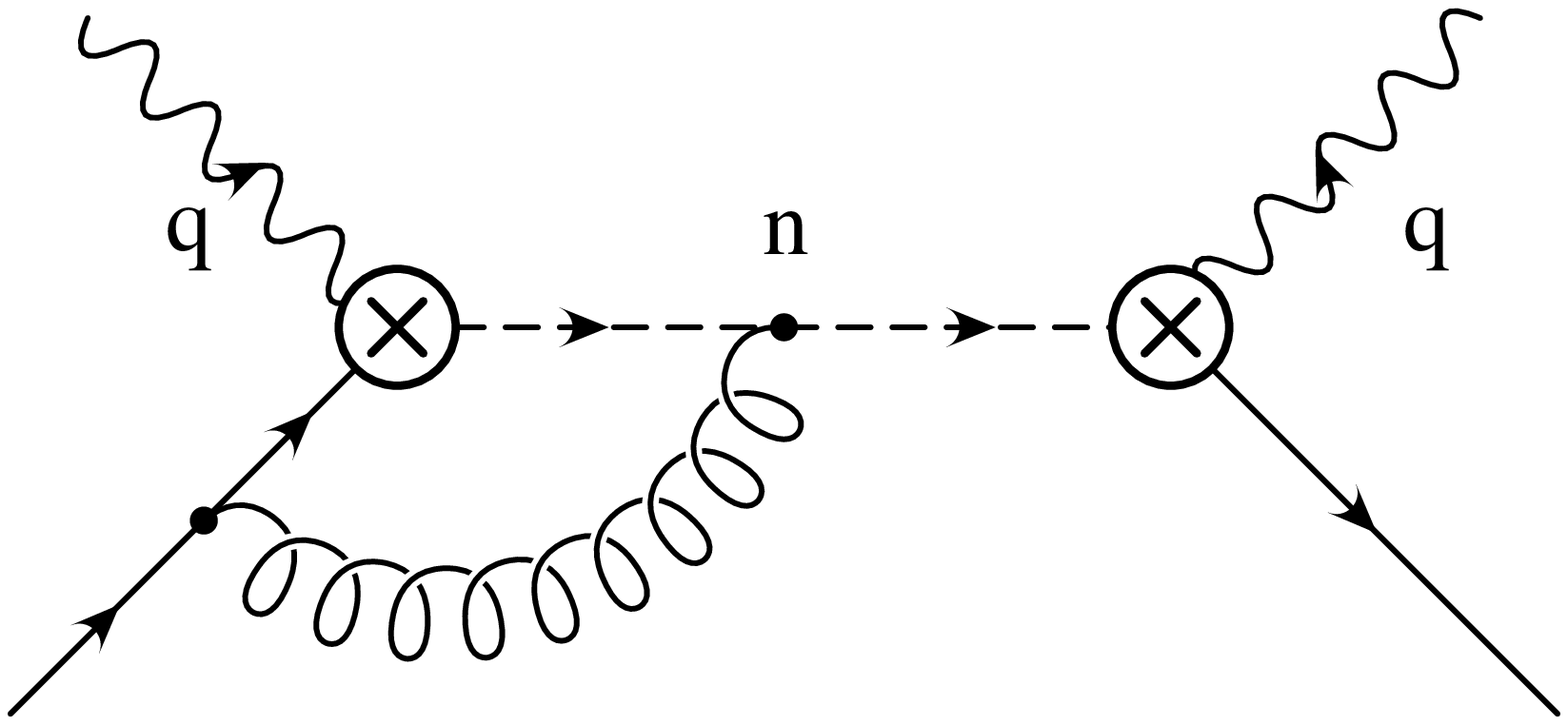} \hfil}\hbox to \size {\hfil(b)\hfil}}
\vbox{\hbox to \size {\hfil \includegraphics[width=4cm]{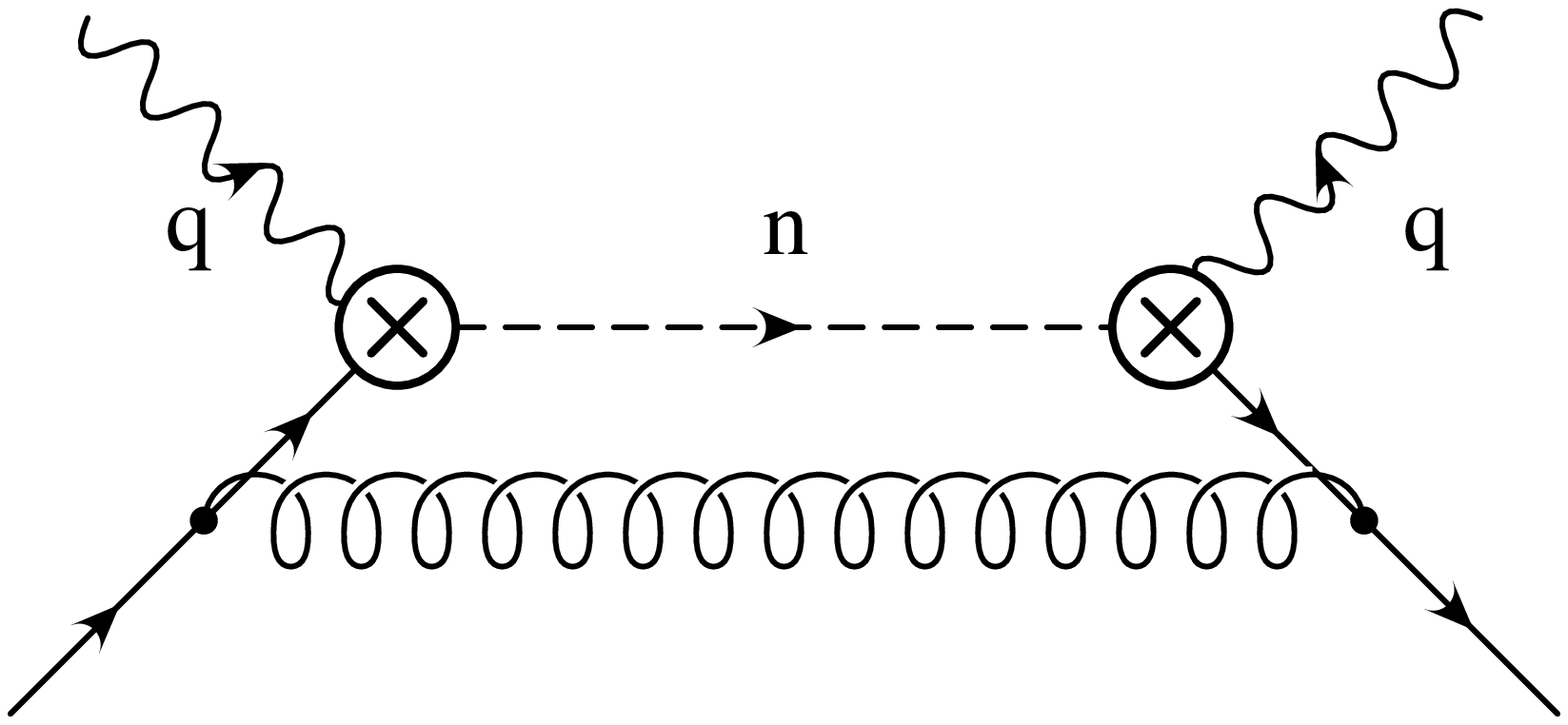} \hfil}\hbox to \size {\hfil(c)\hfil}}}
\hbox{\vbox{\hbox to \size {\hfil \includegraphics[width=4cm]{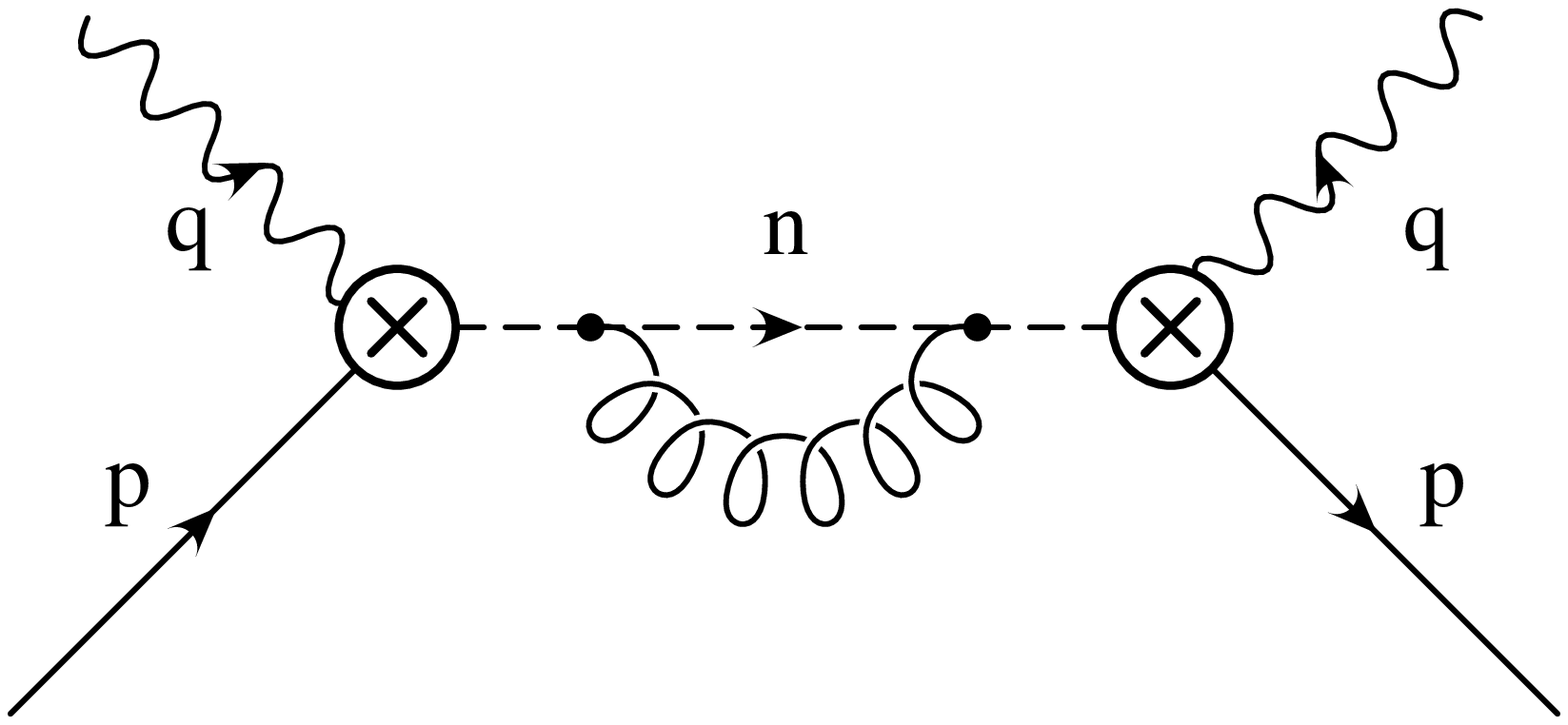} \hfil }\hbox to \size {\hfil(d)\hfil}}
\vbox{\hbox to \size {\hfil \includegraphics[width=4cm]{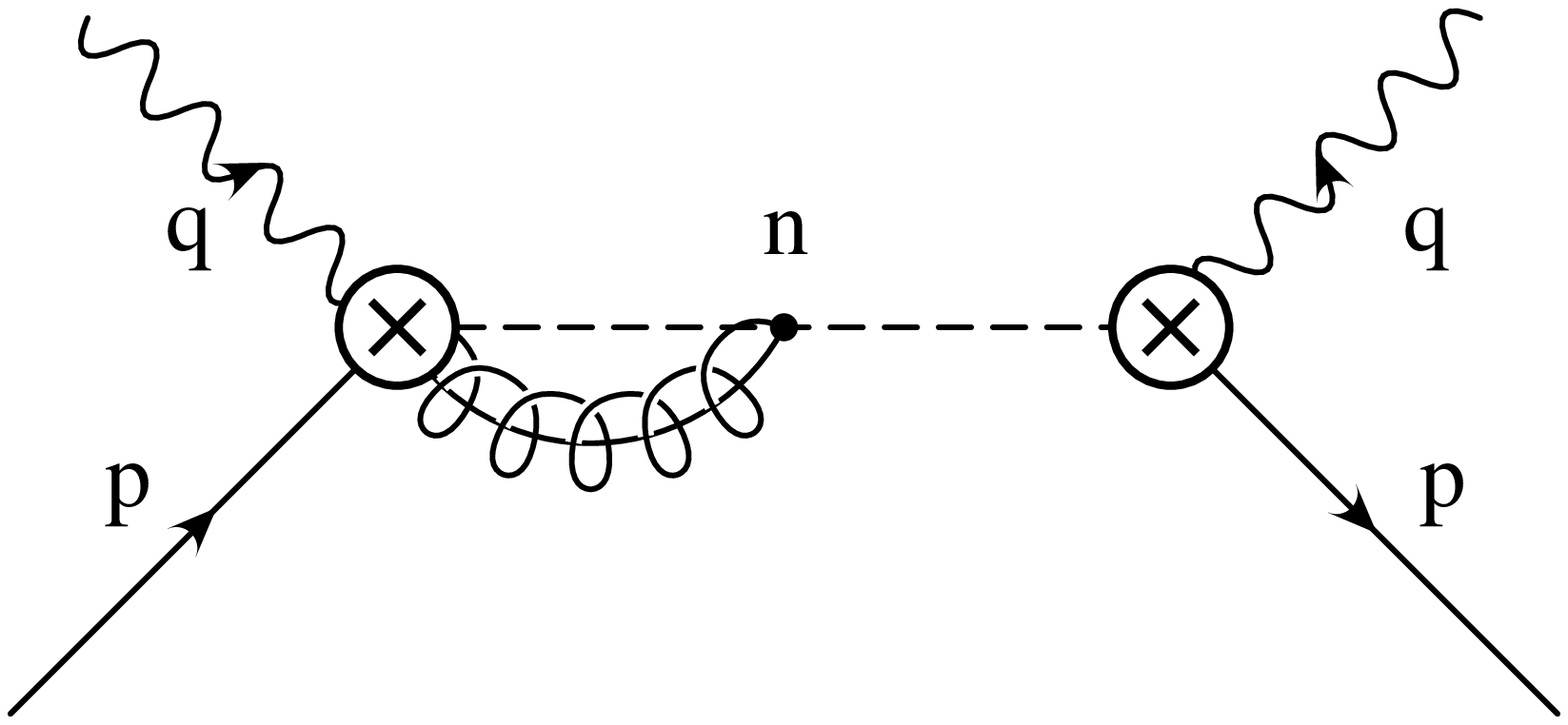} \hfil}\hbox to \size {\hfil(e)\hfil}}
\vbox{\hbox to \size {\hfil \includegraphics[width=4cm]{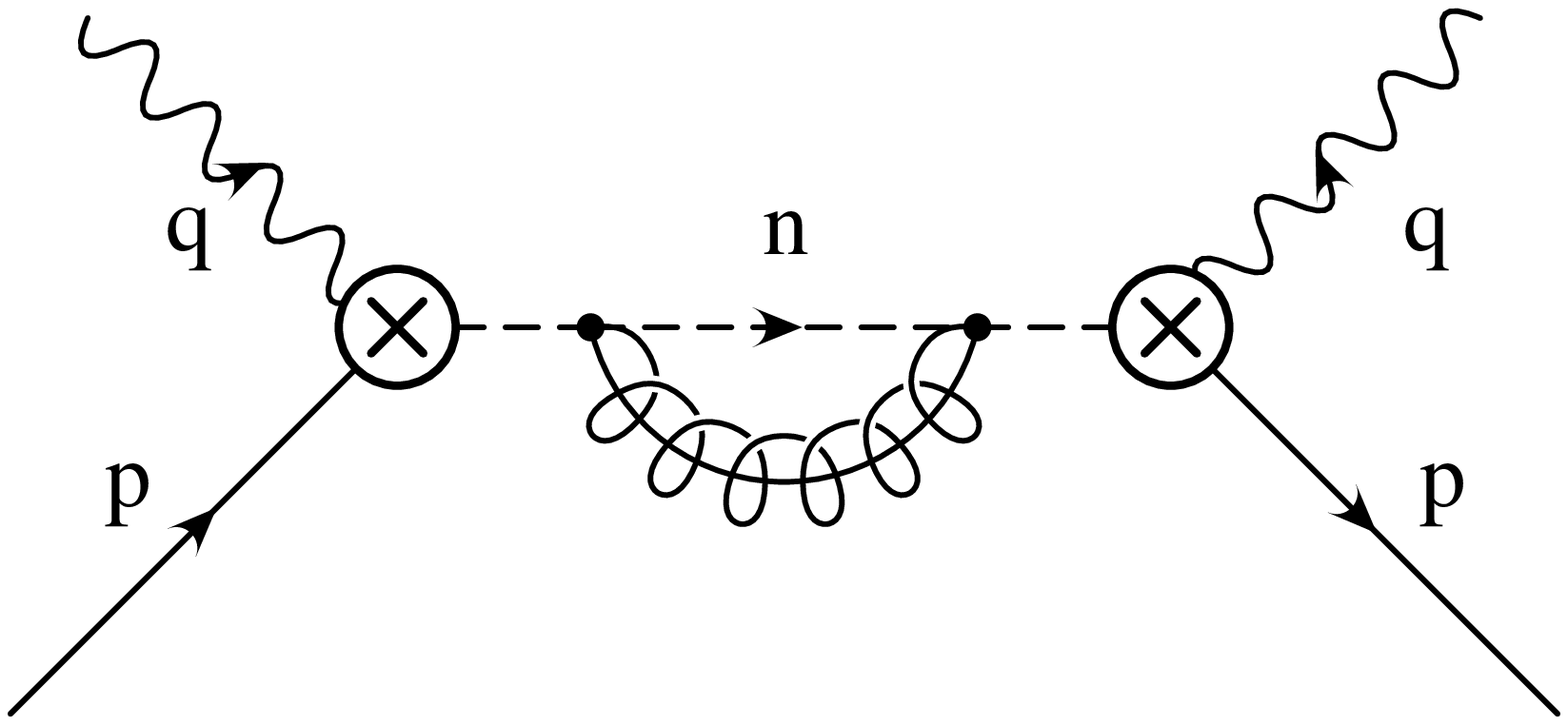} \hfil}\hbox to \size {\hfil(f)\hfil}}}
\hbox{\vbox{\hbox to \size {\hfil}}\vbox{\hbox to \size {\hfil \includegraphics[width=4cm]{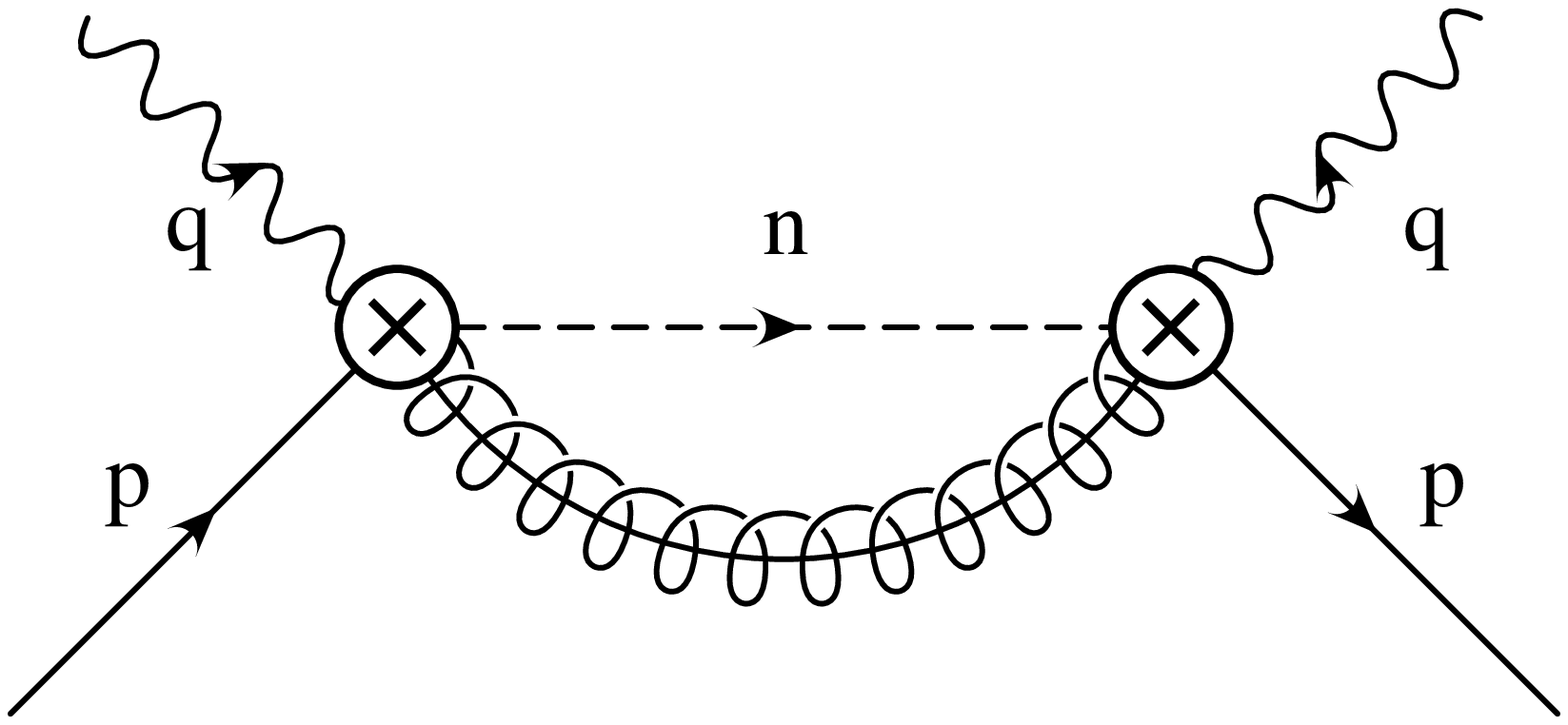} \hfil }\hbox to \size {\hfil(g)\hfil}}}
\caption{One loop correction to the electromagnetic current product in the target rest frame. Graphs (a), (b) and (e) also have mirror image graphs where the gluon attaches to the other side. \label{fig:4}}
\end{figure*}

The one-loop graphs for the matrix element of the current product are shown in Figs.~\ref{fig:4}.
Graphs \ref{fig:4}(a--d) have the same value in the theories above and below $Q^2(1-x)$, since the interaction of an ultrasoft gluon with a collinear quark is the same as the vertex generated by the Wilson line $Y$ in the operator $O_q$, and the external ultrasoft quark fields are unchanged at the matching scale. The matching correction is given by \ref{fig:4}(e--g); these graphs are absent below $Q^2(1-x)$ since the $n$-collinear modes have been integrated out.

The collinear graph Fig.~\ref{fig:4}(e) gives for the spin-averaged matrix element
\begin{eqnarray}
I_{c,1}&=&\text{Disc}\ {g^2 \over 2 \pi}  \int { {\rm d}^d k \over (2 \pi )^d} 
\text{Tr} \frac12 \xslash p \gamma^\nu  { \xslash{n} \bar n \cdot(p+q) \over 2 (p + q )^2}\nn
&&  \times {\xslash{\bar n}\ n^\alpha\over 2}{ \xslash{n} \bar n \cdot(p+q-k) \over 2 (p+q - k )^2}  \gamma^\mu
{  1 \over -\bar n \cdot k}
\bar n_\alpha {1 \over k^2}\nn
&=&\text{Disc}\ {g^2 \over 2 \pi}T^{\mu \nu} p^+ \int { {\rm d}^d k \over (2 \pi )^d} 
  {  \bar n \cdot(p+q) \over (p +q )^2}\nn
&&\times  {  \bar n \cdot(p+q-k) \over  (p+q - k )^2}  
{  1 \over -\bar n \cdot k}
 {1 \over k^2} \ .
\label{6.09}
\end{eqnarray}
Comparing with Eq.~(\ref{5.05}) gives
\begin{eqnarray}
I_{c,1}&=&\text{Disc}\ {i \over 2 \pi}\left({-g^2 \over 8 \pi^2}\right) p^+ T^{\mu \nu} {  \bar n \cdot(p+q) \over (p +q )^2} \nn
&&  \times \Biggl[-{1\over \epsilon^2} - {1 \over \epsilon} + {1\over \epsilon} \ln {-(p+q)^2 \over \mu^2} - \frac 12 \ln^2 {-(p+q)^2 \over \mu^2} \nn
&& + \ln {-(p+q)^2 \over \mu^2}-2 + {\pi^2 \over 12} \Biggr] \nn
&\approx& \text{Disc}\ {i \over 2 \pi}\left({-g^2 \over 8 \pi^2}\right) T^{\mu \nu} { 1 \over (1-y) +i 0^+ }\nn
&&  \times \Bigl[-{1\over \epsilon^2} - {1 \over \epsilon} + {1\over \epsilon} \ln {Q^2(y-1)-i0^+  \over y \mu^2}  \nn
&& - \frac 12 \ln^2 {Q^2(y-1)-i0^+  \over y \mu^2} + \ln  {Q^2(y-1)-i0^+  \over y \mu^2} \nn
&& -2 + {\pi^2 \over 12} \Biggr] \theta(y),
\label{6.09a}
\end{eqnarray}
where $y \ge 0$ since $q^+ < 0$, and $q^- = -Q^2/q^+ >0$. The kinematic region where $q^- <0$ is infinitely far away in the effective theory, and is described by the effective theory with a different value of the label momentum. The graph with gluon attached to the other vertex gives the same contribution, Eq.~(\ref{6.09a}).

The collinear graph Fig.~\ref{fig:4}(f) is given by the tree diagram, times the negative of the wavefunction diagram Eq.~(\ref{4.08}) evaluated with $p^2 \to (p+q)^2$, and gives
\begin{eqnarray}
I_{c,2}&=& \text{Disc}\ {i\over 2 \pi}\left(-{g^2 \over 8 \pi^2} \right) T^{\mu \nu} {p^+ (p^-+q^-) \over (p+q)^2 + i 0^+}
\nn
&& \times  \Biggl[ {1\over 2 \epsilon}
+ \frac 1 2 - \frac 1 2 \ln {-(p+q)^2 \over \mu^2} \Bigr] \nn
&\approx& \text{Disc}\ {i\over 2 \pi}\left(-{g^2 \over 8 \pi^2} \right) T^{\mu \nu} {1 \over  (1-y)  + i 0^+} \nn
&& \times  \Bigl[ {1\over 2 \epsilon}
+ \frac 1 2 - \frac 1 2 \ln {Q^2(y-1) -i0^+ \over y \mu^2} \Biggr]\theta(y) .
\label{6.11}
\end{eqnarray}

The graph Fig.~\ref{fig:4}(g) vanishes, since the collinear gluon emission vertex is proportional to $\bar n^\alpha$, and $\bar n^2=0$.

The total collinear contribution is the sum of twice Eqs.~(\ref{6.09}) and Eq.~(\ref{6.11}),
\begin{eqnarray}
I_c&=& \text{Disc}\ {i \over 2 \pi}\left({-g^2 \over 8 \pi^2}\right) T^{\mu \nu} { 1 \over (1-y) +i 0^+ }\nn
&&\times  \Biggl[-{2\over \epsilon^2} - {3 \over 2 \epsilon} + {2\over \epsilon} \ln {Q^2(y-1) -i0^+ \over y \mu^2}  \nn
&& -   \ln^2 {Q^2(y-1)-i0^+  \over y \mu^2}  + \frac 3 2 \ln  {Q^2(y-1)  -i0^+\over y \mu^2} \nn
&& - \frac 7 2  + {\pi^2 \over 6} \Biggr] .
\label{6.12}
\end{eqnarray}
The collinear counterterms cancel the $1/\epsilon$ terms, so that the remaining matching contribution is finite. The collinear counterterm contribution $(\ln -p^2)/\epsilon=[\ln Q^2(y-1)]/\epsilon $ is no longer infrared sensitive, since the scale $Q^2(1-y)$ is now considered a large scale.

The discontinuity of the remaining terms can be written in terms of $+$ distributions.  The distribution $1/(1-y)_+$ is defined by
\begin{eqnarray}
\int_0^1 {\rm d}y f(y) {1 \over (1-y)_+} &\equiv & \int_0^1 {\rm d}y  {f(y)-f(1) \over (1-y)_+},
\label{6.13}
\end{eqnarray}
so that
\begin{eqnarray}
\int_0^1 {\rm d}y  {1 \over (1-y)_+} &=& 0.
\label{6.14}
\end{eqnarray}

The discontinuity of
\begin{eqnarray}
{i \over 2 \pi}  {\ln (y - 1 - i \eta) \over 1 - y + i \eta} \theta(y)
\label{6.15}
\end{eqnarray}
is given by the difference of the expression for $\eta \to 0^+$ and $\eta \to 0^-$,
\begin{eqnarray}
\text{Disc}\ {i \over 2 \pi}  {\ln (y - 1 - i \eta) \over 1 - y + i \eta}\theta(y) = {1 \over 1 - y}\theta(0 \le y < 1).
\label{6.16}
\end{eqnarray}
The singular terms at $y=1$ can be obtained by integrating Eq.~(\ref{6.13}) over $0 \le y \le \Lambda$, where $\Lambda > 1$. Then
\begin{eqnarray}
{i \over 2 \pi} \int_0^\Lambda {\rm d}y {\ln (y - 1 - i \eta) \over 1 - y + i \eta}
&=& - {i \over 4 \pi}  \left[ \ln^2 \Lambda - (- i \pi)^2 \right], \nn
\label{6.17}
\end{eqnarray}
irrespective of the sign of $\eta$. This has no discontinuous part, so the integral of the discontinuity is zero. This gives
\begin{eqnarray}
\text{Disc}\ {i \over 2 \pi}  {\ln (y - 1 - i \eta) \over 1 - y + i \eta}\theta(y) = {1 \over (1 - y)_+} .
\label{6.18}
\end{eqnarray}
Similarly,
\begin{eqnarray}
{i \over 2 \pi} \int_0^\Lambda {\rm d}y {\ln^2 (y - 1 - i \eta) \over 1 - y + i \eta}
&=& - {i \over 6 \pi} \left[ \ln^3 \Lambda - (- i \pi)^3 \text{sign}\, \eta \right],\nn
\label{6.19}
\end{eqnarray}
so that
\begin{eqnarray}
\text{Disc}\ {i \over 2 \pi} \int_0^\Lambda {\rm d}y {\ln^2 (y - 1 - i \eta) \over 1 - y + i \eta}
&=& - {\pi^2 \over 3},
\label{6.20}
\end{eqnarray}
and
\begin{eqnarray}
\text{Disc}\ {i \over 2 \pi} {\ln^2 (y - 1 - i \eta) \over 1 - y + i \eta} \theta(y)
&=& - {\pi^2 \over 3}\delta(1-y)  \nn
&& + \left[ {2 \ln (1-y) \over 1 - y } \right]_+  .
\label{6.21}
\end{eqnarray}

Using the above results gives for Eq.~(\ref{6.12}),
\begin{eqnarray}
\text{Disc} &\equiv& T^{\mu \nu} \mathcal{M}(y), \nn
\mathcal{M}(y) &=& {\alpha_s \over 2 \pi } \theta(0 \le y \le 1) \Biggl\{2  {\left[ \ln(1-y) \over (1-y)\right]_+ }  \nn
&& + \Bigl[2  \ln {Q^2\over  \mu^2} 
 - \frac 32  \Bigr]{ 1 \over (1-y)_+ }\nn
 && +
 \Bigl[    \ln^2 {Q^2  \over  \mu^2} 
- \frac 32  \ln  {Q^2  \over  \mu^2}+  \frac 7 2 - {\pi^2 \over 2} \Bigr] \delta(1-y) \Biggr\} , \nn
\label{6.22}
\end{eqnarray}
which defines the matching function $\mathcal{M}(y)$. Formally, $1-y \sim \lambda$ is the expansion parameter so $\ln y = \ln(1 + (1-y)) \to 0$, which was used to simplify Eq.~(\ref{6.22}). The matching condition at one-loop is therefore
\begin{eqnarray}
W^{\mu \nu} &=& T^{\mu \nu} \int {{\rm d}r^+ \over r^+}\left[ \delta\left(1-{-q^+ \over r^+} \right) + \mathcal{M}\left({-q^+ \over r^+} \right) \right] O_q(r^+) .\nn
\label{6.23}
\end{eqnarray}

The moments of this expression are (see Appendix~\ref{app:mom})
\begin{eqnarray}
M_N(\mathcal{M}) &=& {\alpha_s \over 2 \pi } \Biggl[ \left(\frac 3 2 - 2  \ln {Q^2\over  \mu^2}\right) \sum_{j=1}^{N-1} {1 \over j} +2 \sum_{j=1}^{N-1} {H_j \over j}\nn
&& +  \ln^2 {Q^2  \over  \mu^2}  - \frac 32  \ln  {Q^2  \over  \mu^2}+ \frac 7 2 - {\pi^2 \over 2}\Biggr].
\label{6.24}
\end{eqnarray}
which as $N \to \infty$ is
\begin{eqnarray}
M_N(\mathcal{M})  &\to& {\alpha_s \over 2 \pi }C_F \Biggl[  \left(\frac 3 2 - 2  \ln {Q^2\over  \mu^2}\right) \left( \ln N + \gamma_E \right)\nn
&& +  \left( \ln N + \gamma_E \right)^2 +{\pi^2 \over 6} +  \ln^2 {Q^2  \over  \mu^2} \nn
&& - \frac 32  \ln  {Q^2  \over  \mu^2}+ \frac 7 2 - {\pi^2 \over 2}\Biggr].
\label{6.25}
\end{eqnarray}
We have dropped terms of order $1-y$ times the terms retained in Eq.~(\ref{6.22}). All such terms have moments which vanish as $N \to \infty$, so Eq.~(\ref{6.25}) is valid up to corrections which vanish as $N \to \infty$. Letting 
\begin{eqnarray}
\bN = N e^{\gamma_E},
\label{6.26}
\end{eqnarray}
one finds
\begin{eqnarray}
M_N(\mathcal{M})  &\to& {\alpha_s(\mu) \over 2 \pi }C_F \Biggl[ \ln^2 {Q^2  \over \bN \mu^2}  - \frac 32  \ln  {Q^2  \over \bN \mu^2} + \frac 7 2  - {\pi^2 \over 3} \Biggr] .\nn
\label{6.27}
\end{eqnarray}
The logarithms are minimized if $\mu = Q^2/\bN$, so one can match from currents onto the bilocal operator at this scale using
\begin{eqnarray}
M_N(\mathcal{M})  &=& {\alpha_s \left( Q/ \sqrt{\bN} \right) \over 2 \pi }C_F \left[  \frac 7 2  - {\pi^2 \over 3} \right] .
\label{6.28}
\end{eqnarray}
The matching condition has no large logarithms, and so does not depend on $\bN$ in the $\bN \to \infty$ limit, except through the argument of $\alpha_s$.

\subsection{Breit Frame}

\begin{figure}
\includegraphics[width=4cm]{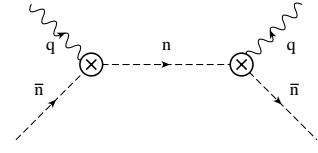}
\caption{Tree graph for the product of currents in the Breit frame. \label{fig:5}}
\end{figure}

The computation of the matching condition in the Breit frame is similar to that in the target rest frame. The tree-level graph is given in Fig.~\ref{fig:5}.
It gives the matching relation Eq.~(\ref{6.08}) where the quark distribution operator in the effective theory is now
\begin{eqnarray}
O_q(r^+) \!&=&\! {1 \over 4 \pi} \int_{-\infty}^\infty\!\!\! \diff z e^{-i z r^+}
\left[ \bar \xi_{\bar n}W_{\bar n}\right]\!\! (n z)  \xslash nY_n(z,0) \left[ W^\dagger_{\bar n}   \xi_{\bar n}\right] \!\! \left( 0 \right) \nn
\label{6.28a}
\end{eqnarray}
which has the same form as Eq.~(\ref{6.04}), except the external fields are $\bn$-collinear quarks instead of ultrasoft quarks. The collinear Wilson lines are required by collinear gauge invariance.

\begin{figure*}
\def\size{5.75 cm}
\hbox{\vbox{\hbox to \size {\hfil \includegraphics[width=4cm]{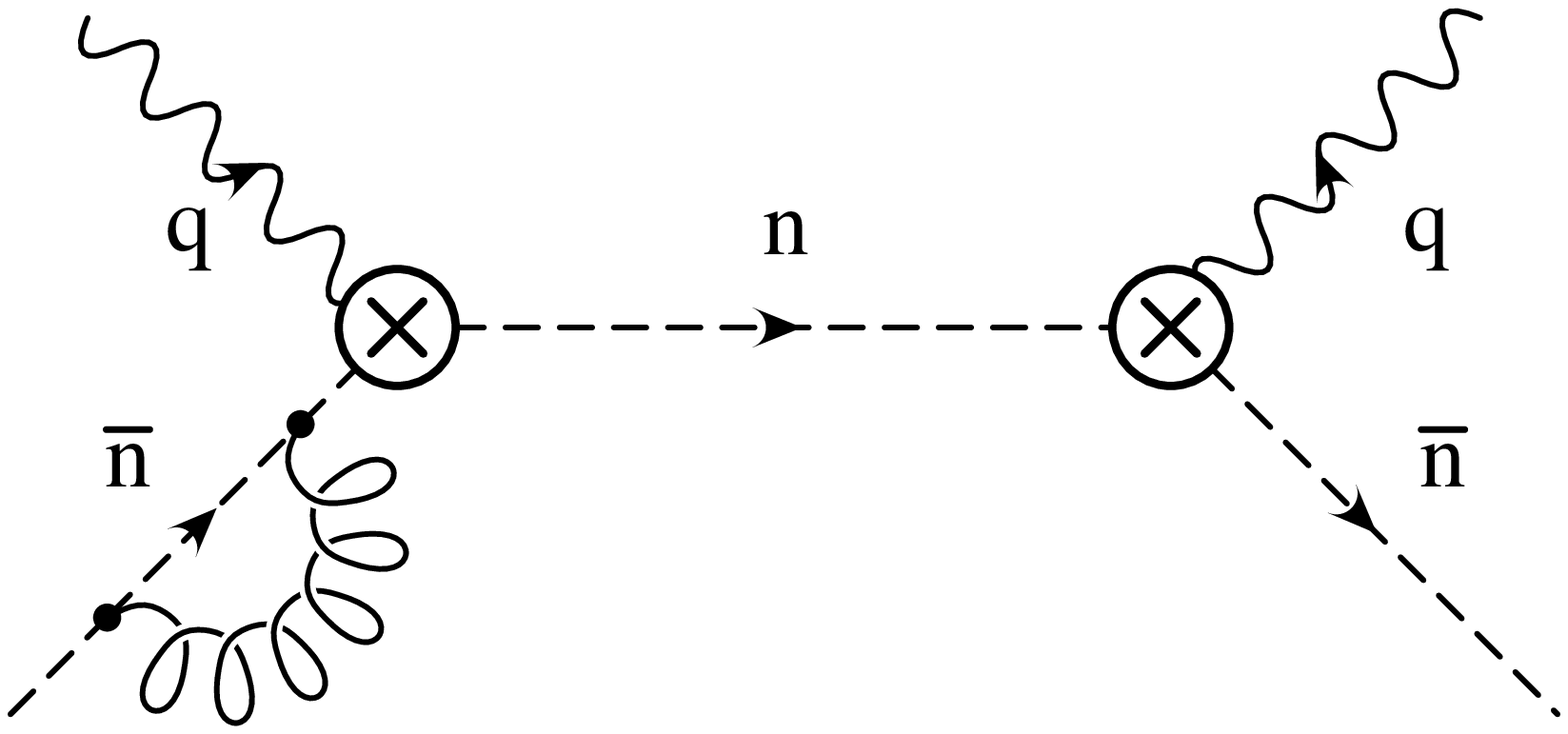} \hfil }\hbox to \size {\hfil(a)\hfil}}
\vbox{\hbox to \size {\hfil \includegraphics[width=4cm]{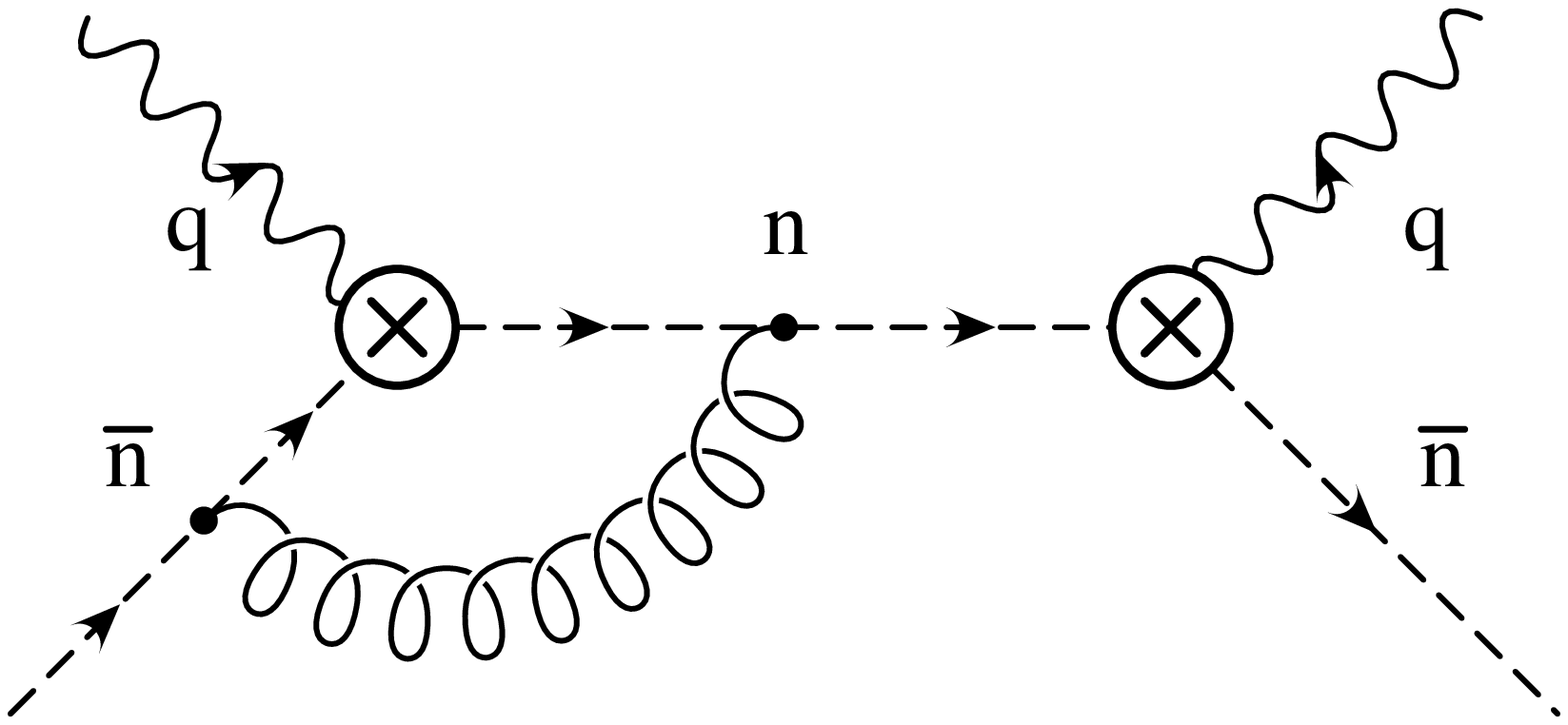} \hfil}\hbox to \size {\hfil(b)\hfil}}
\vbox{\hbox to \size {\hfil \includegraphics[width=4cm]{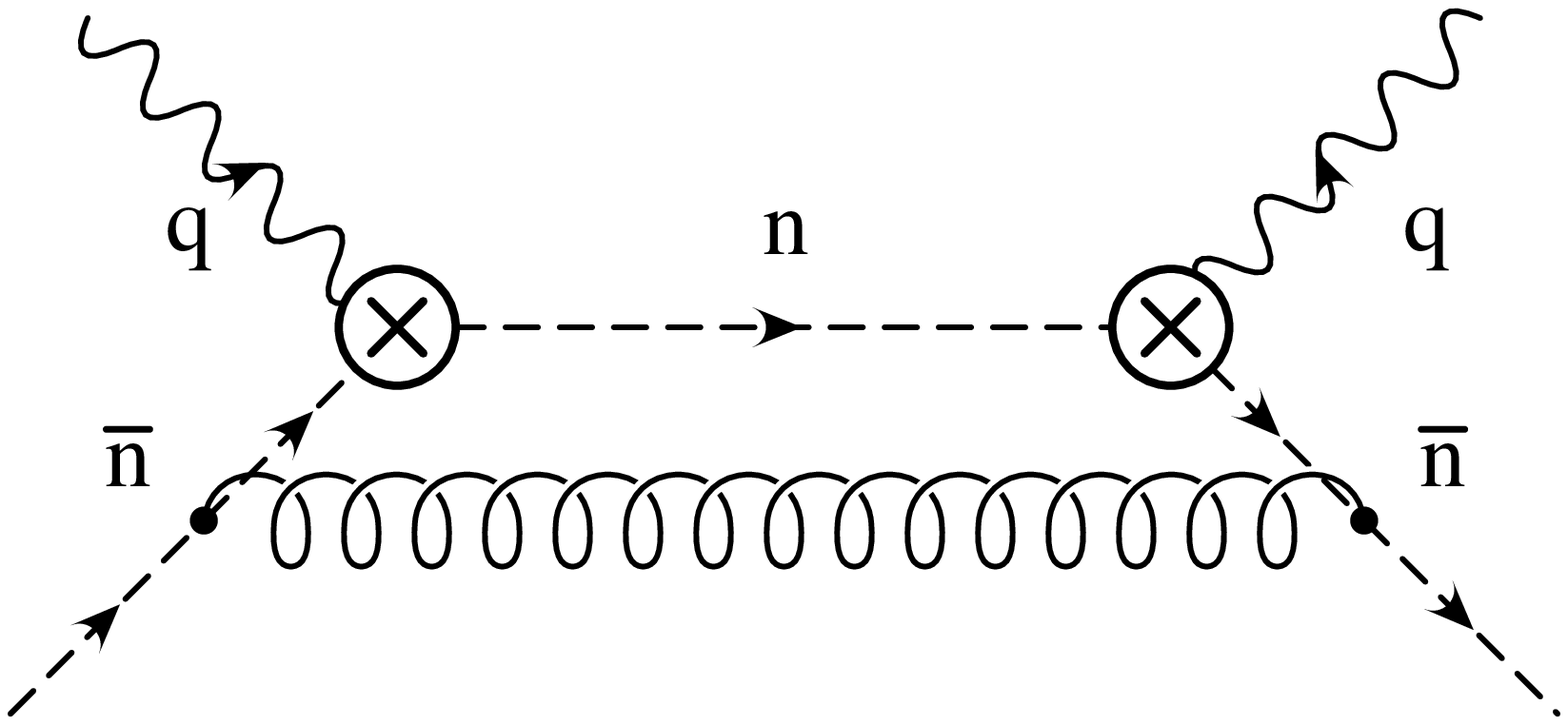} \hfil}\hbox to \size {\hfil(c)\hfil}}}
\hbox{\vbox{\hbox to \size {\hfil \includegraphics[width=4cm]{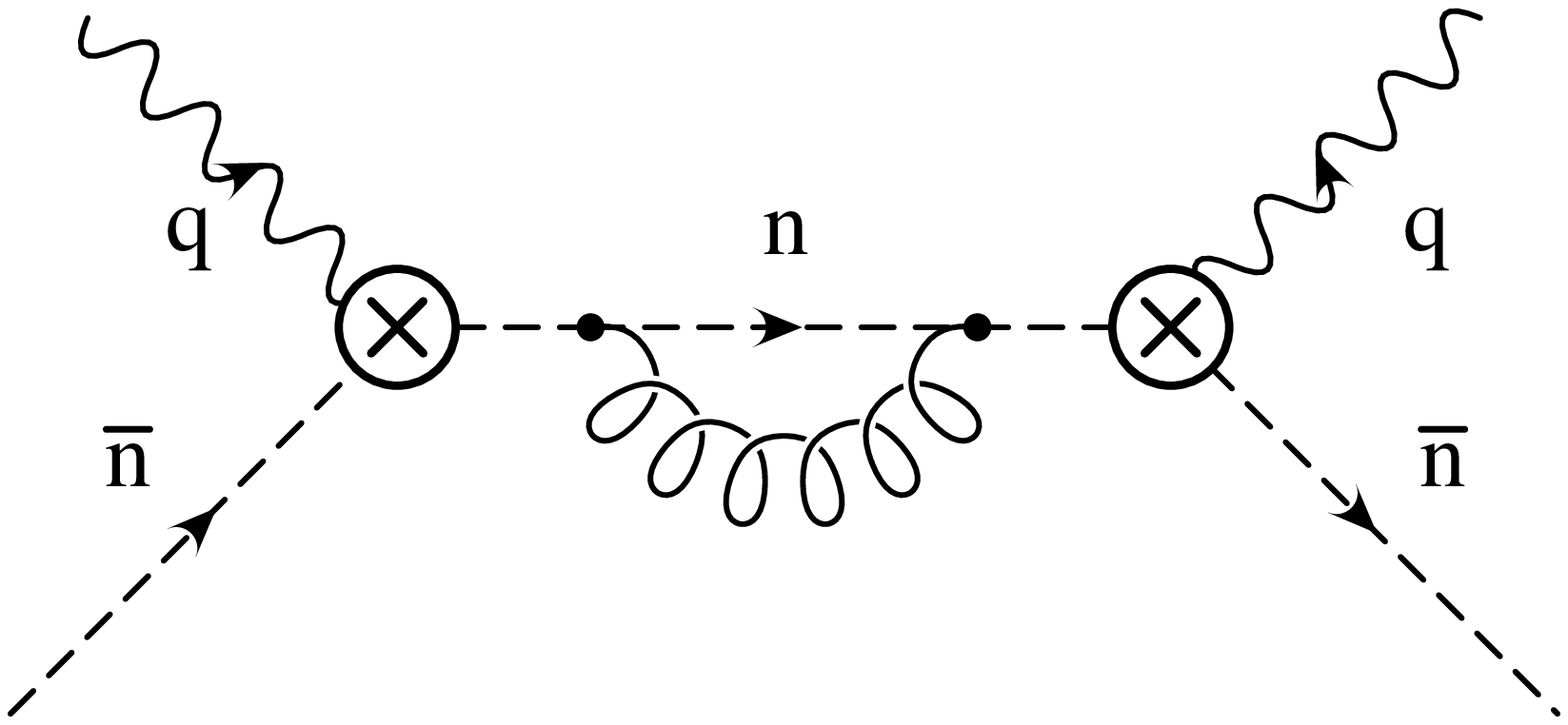} \hfil }\hbox to \size {\hfil(d)\hfil}}
\vbox{\hbox to \size {\hfil \includegraphics[width=4cm]{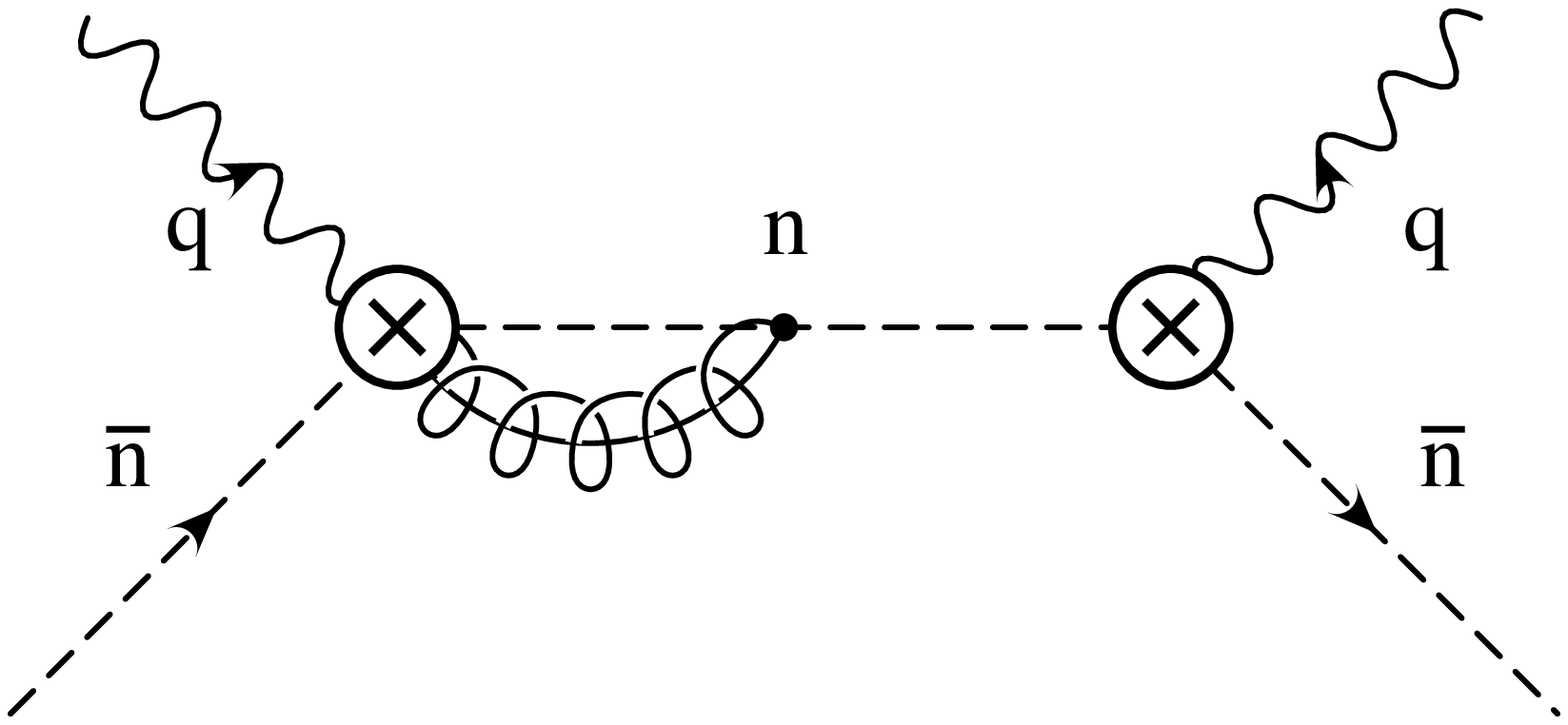} \hfil}\hbox to \size {\hfil(e)\hfil}}
\vbox{\hbox to \size {\hfil \includegraphics[width=4cm]{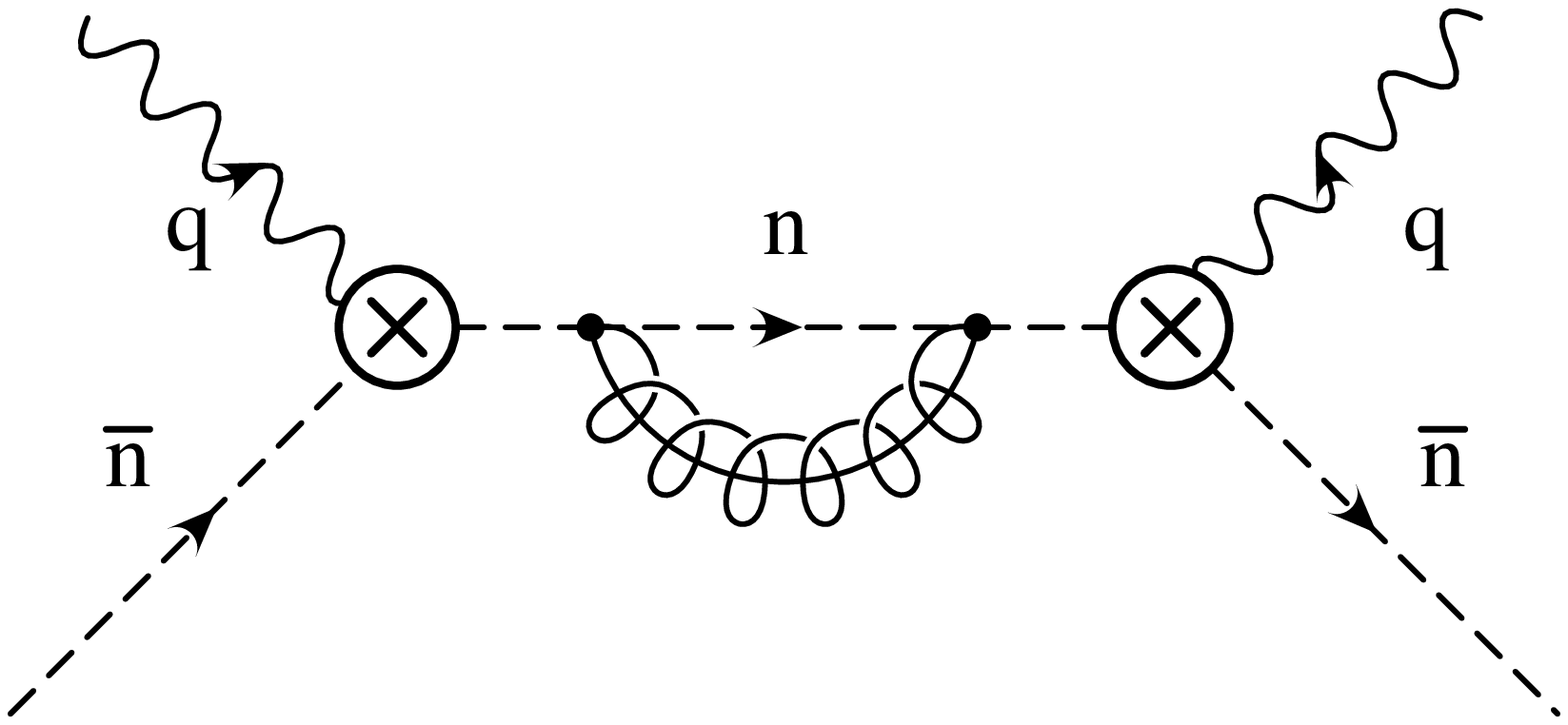} \hfil}\hbox to \size {\hfil(f)\hfil}}}
\hbox{\vbox{\hbox to \size {\hfil \includegraphics[width=4cm]{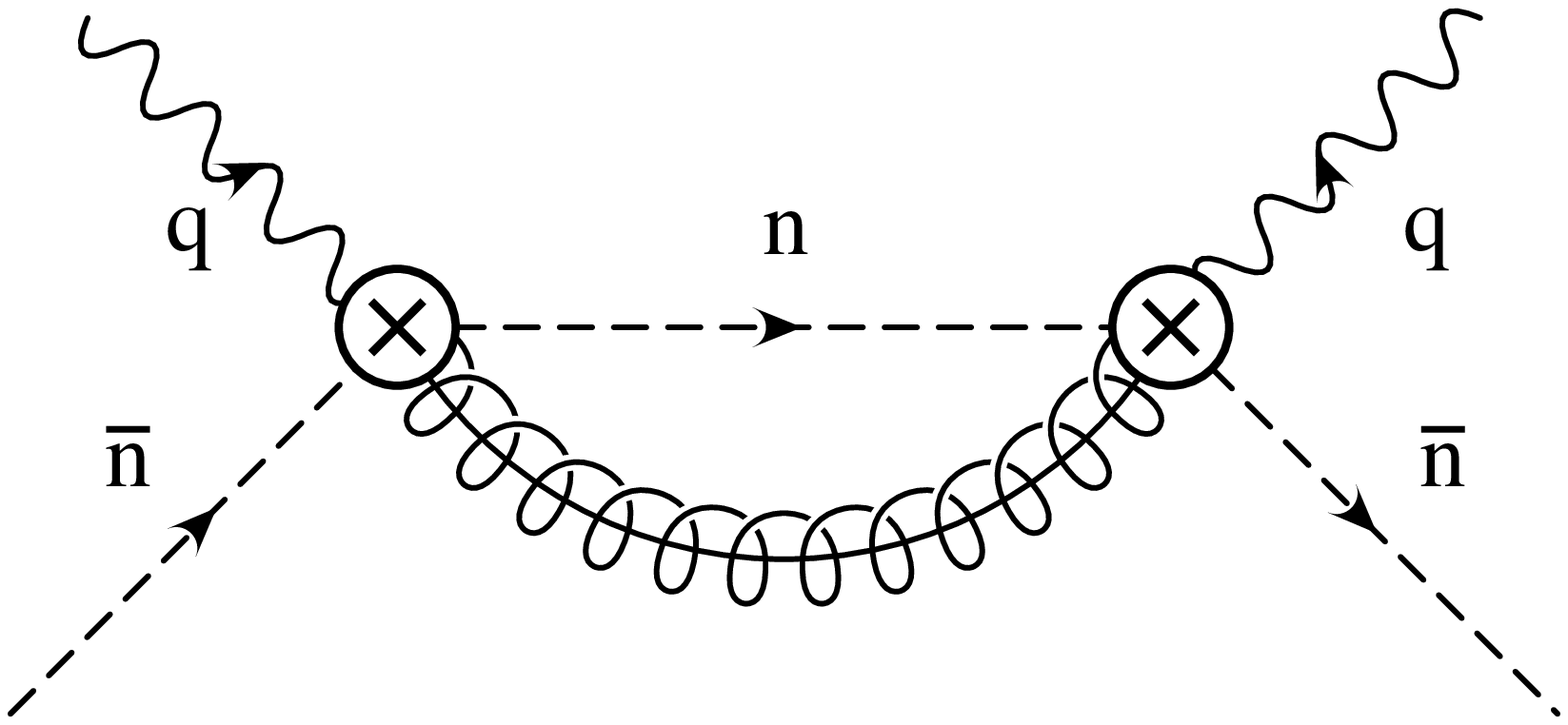} \hfil }\hbox to \size {\hfil(g)\hfil}}
\vbox{\hbox to \size {\hfil \includegraphics[width=4cm]{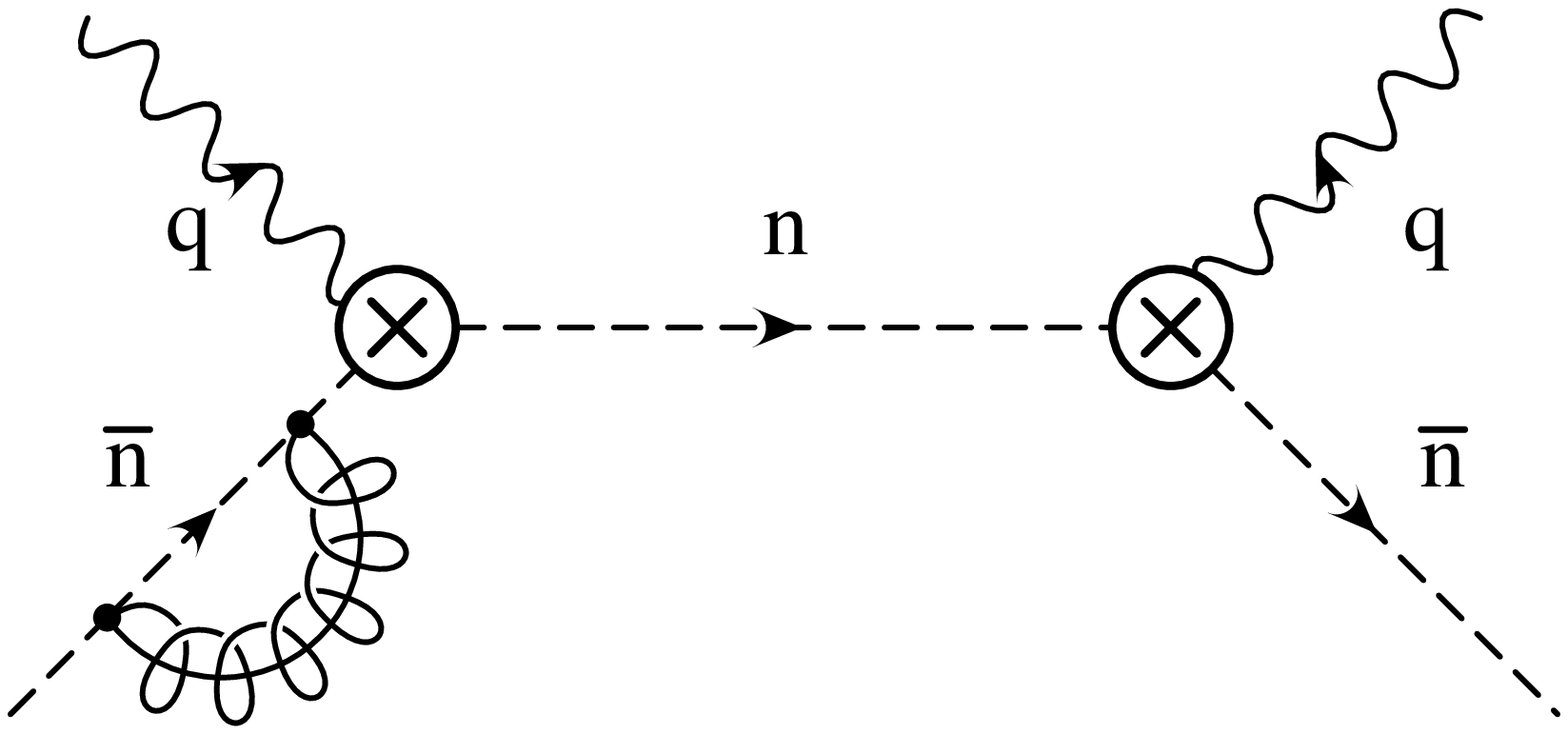} \hfil}\hbox to \size {\hfil(h)\hfil}}
\vbox{\hbox to \size {\hfil \includegraphics[width=4cm]{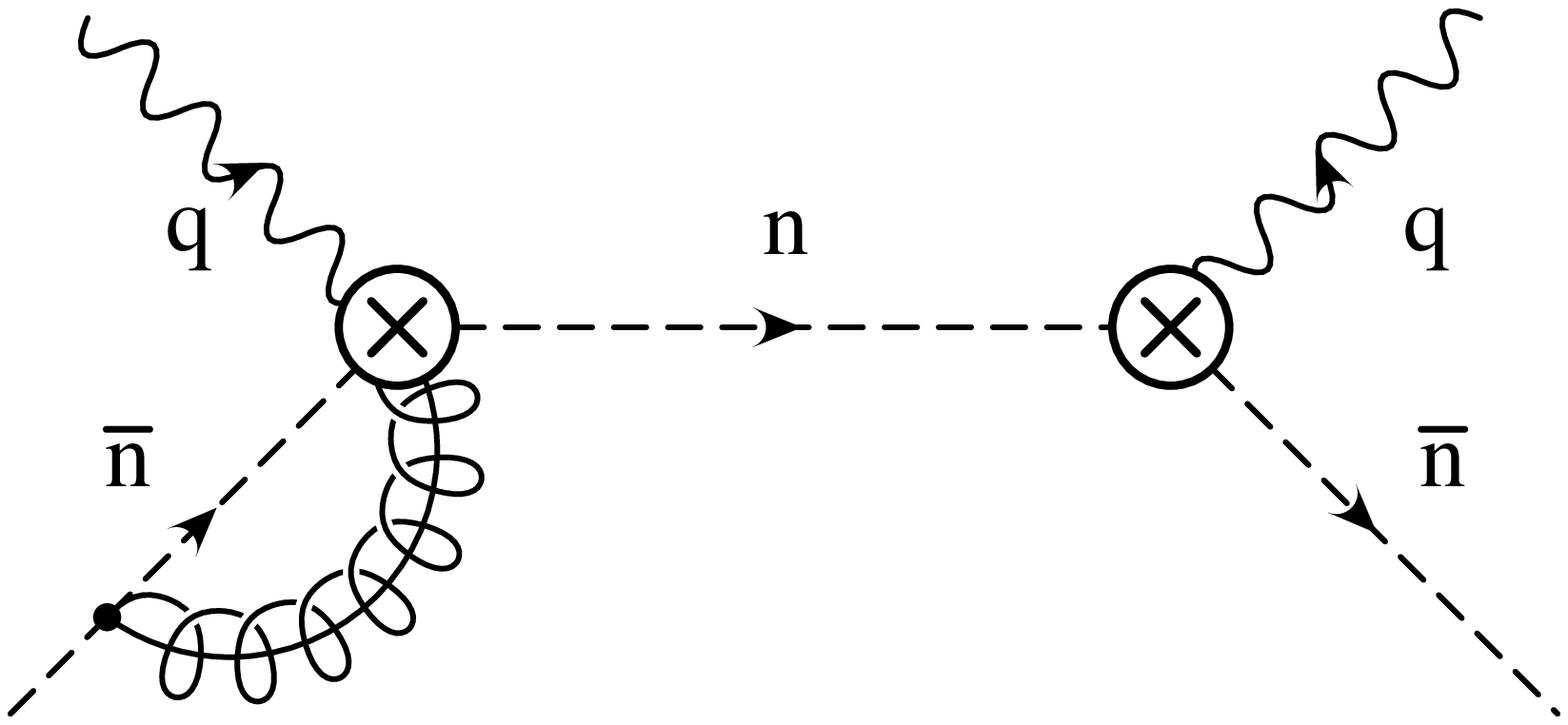} \hfil}\hbox to \size {\hfil(i)\hfil}}}
\hbox{\vbox{\hbox to \size {\hfil}}\vbox{\hbox to \size {\hfil \includegraphics[width=4cm]{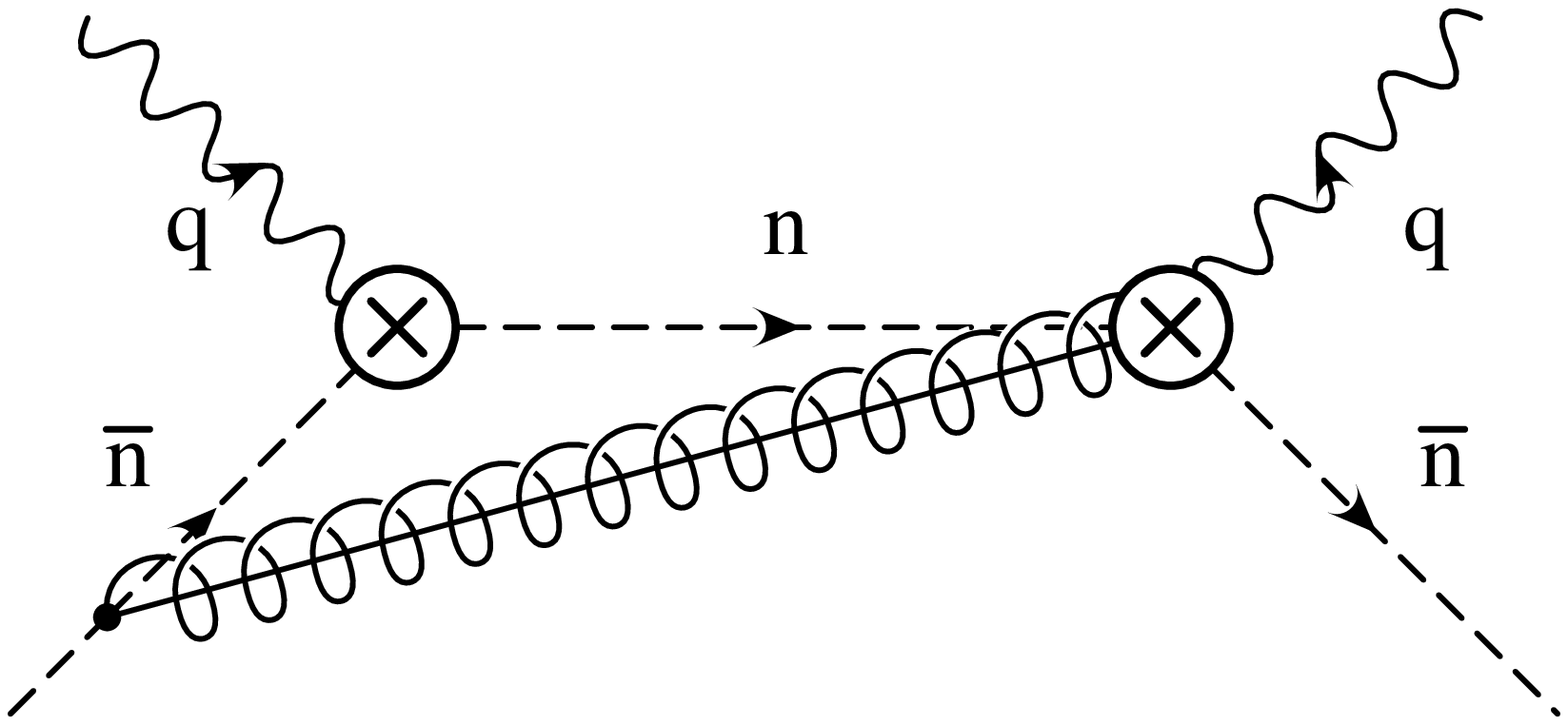} \hfil }\hbox to \size {\hfil(j)\hfil}}}
\caption{One loop correction to the electromagnetic current product in the Breit frame. Graphs (a), (b), (e) and (h--j) also have mirror image graphs where the gluon attaches to the other side. \label{fig:6}}
\end{figure*}

The one-loop graphs in the Breit frame are given in Fig.~\ref{fig:6}.
All the graphs except Figs.~\ref{fig:6}(e--g) are also present in the effective theory, so only these graphs need to be computed for the matching condition. The graphs are identical to the corresponding graphs in the target rest frame, so the matching condition is the same, Eqs.~(\ref{6.22},\ref{6.23}), with the replacement of Eq.~(\ref{6.04}) by Eq.~(\ref{6.28a}).

\section{Renormalization of Parton Distributions \label{sec:run}}

Parton operator renormalization gives the standard DGLAP evolution for parton distribution functions. The computation of the anomalous dimension in the target rest frame is identical to the computation using QCD quarks in Ref.~\cite{Collins}. This calculation is sketched here, so that results can be compared with the computation in the Breit frame. We will only discuss renormalization of quark operators, which gives the evolution of the flavor non-singlet quark distribution. Singlet evolution mixes quark and gluon operators. This complication does not shed additional light on SCET, and will be omitted here.

\subsection{Target Rest Frame}

The renormalization of the (non-singlet) quark operator can be determined by evaluating its matrix element in a free quark state of momentum $p$. The spin-averaged tree-level matrix element of $O_q(w p^+)$ is given by Fig.~\ref{fig:4z}, and is
\begin{eqnarray}
\me{p}{ O_q(w p^+) } {p} &=& \delta(1-w).
\label{7.01}
\end{eqnarray}

\begin{figure*}
\def\size{5.75 cm}
\hbox{\vbox{\hbox to \size {\hfil \includegraphics[width=4cm]{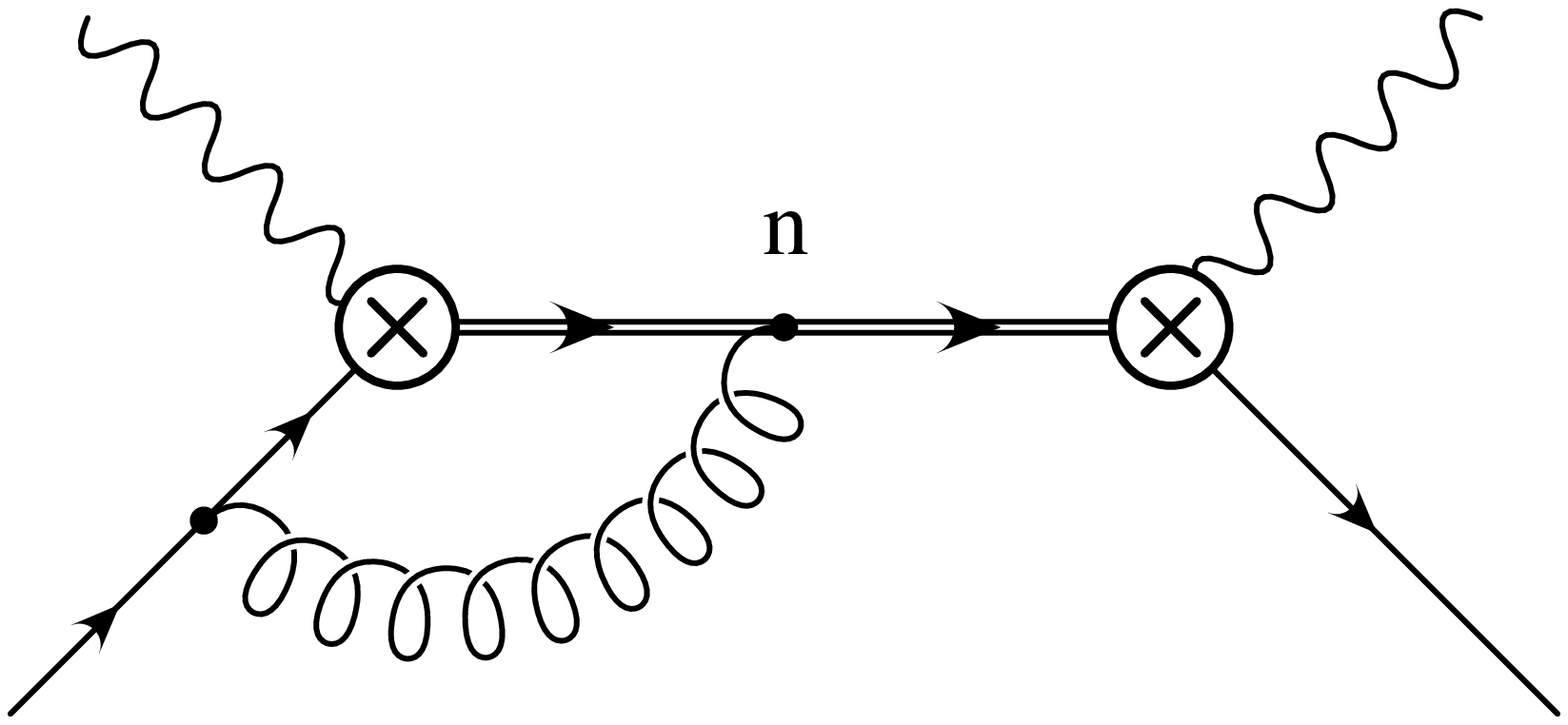} \hfil }\hbox to \size {\hfil(a)\hfil}}
\vbox{\hbox to \size {\hfil \includegraphics[width=4cm]{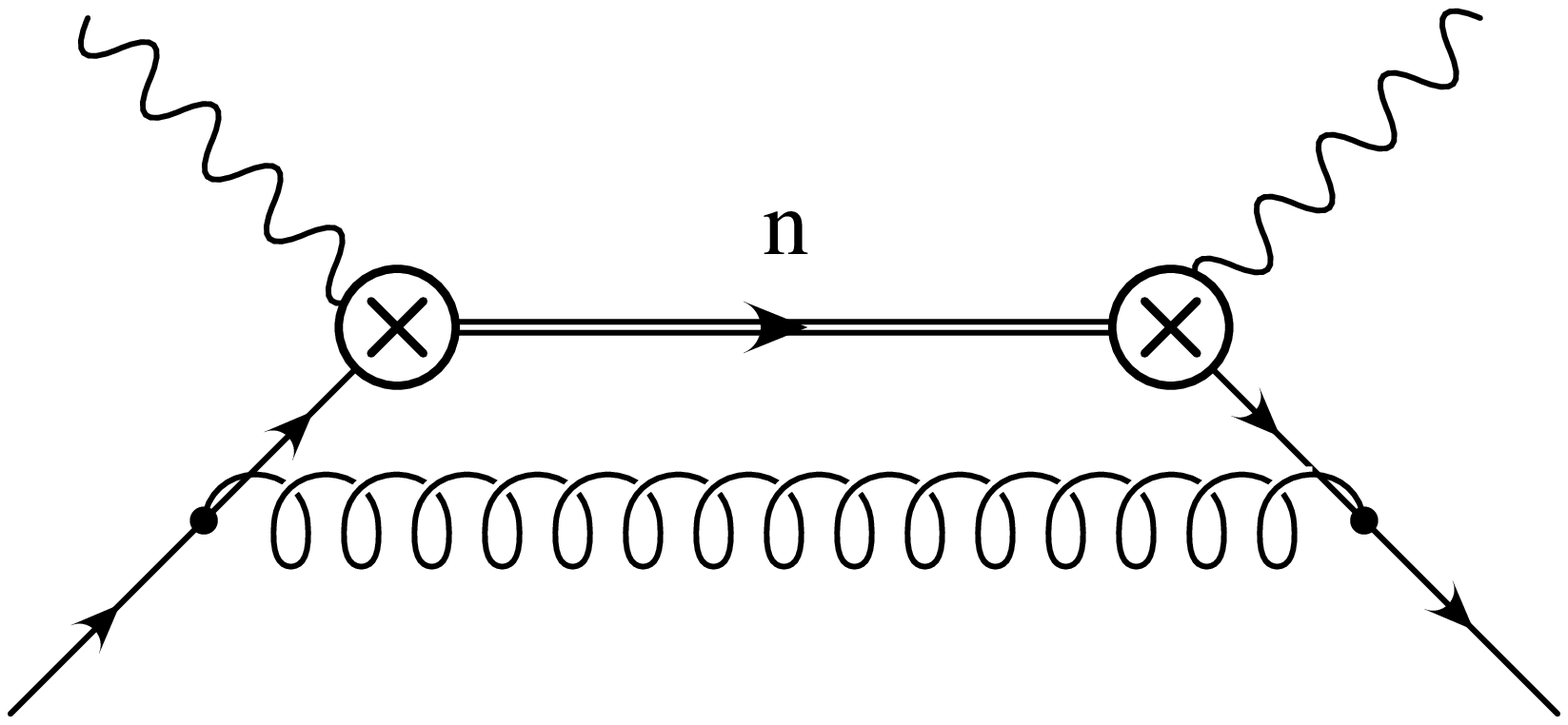} \hfil}\hbox to \size {\hfil(b)\hfil}}
\vbox{\hbox to \size {\hfil \includegraphics[width=4cm]{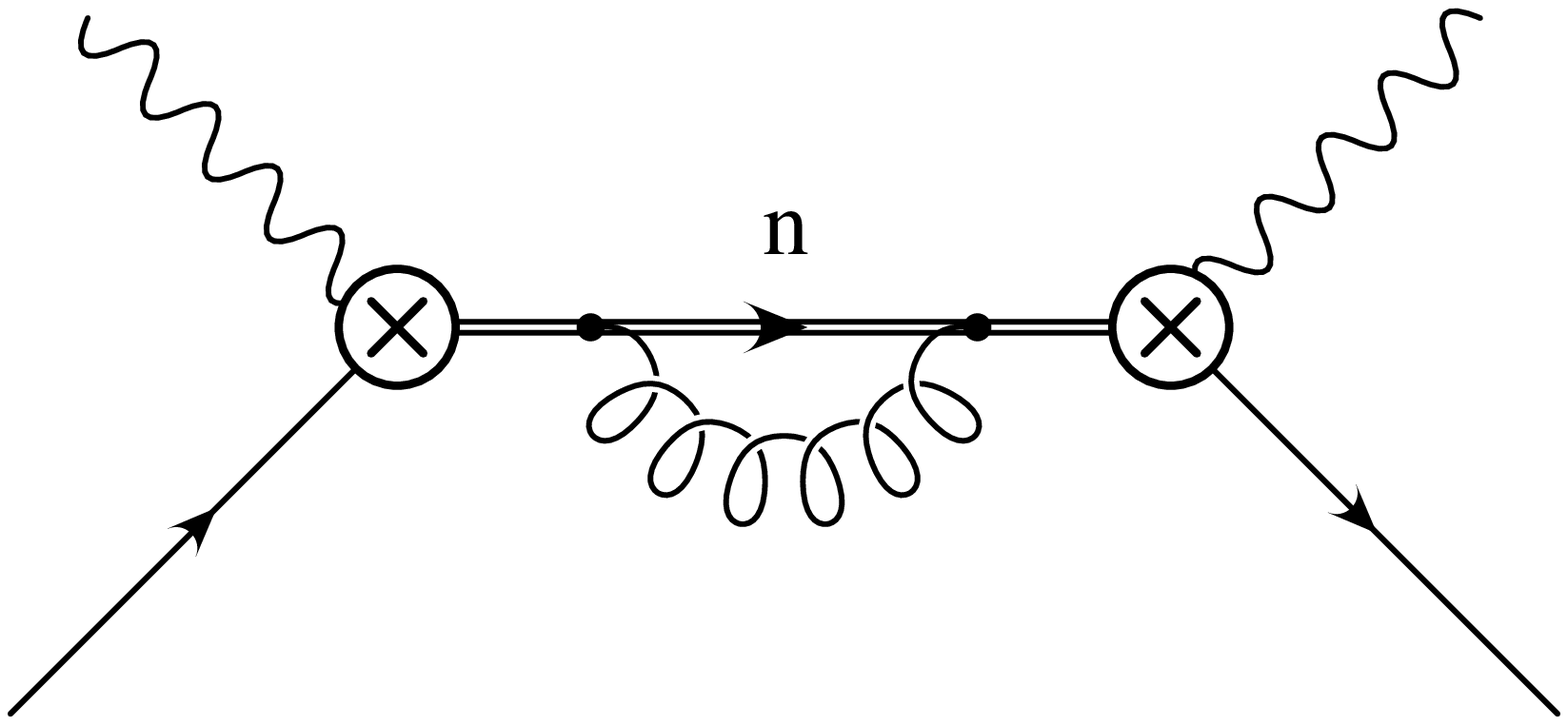} \hfil}\hbox to \size {\hfil(c)\hfil}}
}
\caption{One loop correction to the quark distribution function. The double line is the ultrasoft Wilson line $Y$. \label{fig:7}}
\end{figure*}

The one loop graphs are those of Fig.~\ref{fig:7}, as well as wavefunction graphs.
To compute the anomalous dimensions, we only need the divergent parts of the diagrams. Fig.~\ref{fig:7}(a) is the discontinuity of
\begin{eqnarray}
I_1 &=& {g^2 C_F\over 4 \pi} \int { {\rm d}^d k \over (2 \pi)^d} \frac12 \text{Tr} \ \xslash{p} {1 \over n \cdot \left( p - r \right) +i0^+ } n^\mu \nn
&& {1 \over n \cdot \left( p - k - r \right)+i0^+  } \xslash{n} { \xslash p - \xslash{k} \over (p-k)^2+i0^+} \gamma_\mu {1 \over k^2+i0^+}. \nn
\label{7.02}
\end{eqnarray}
Evaluating the $k^-$ integral by contours, doing the $\mathbf{k}_\perp$ integral, and using the substituting $k^+=zp^+$ gives for the infinite part
\begin{eqnarray}
I_1&=& i{g^2 C_F \over  16\pi^3 \epsilon}\int_0^{1} {\rm d}z { \left(1-z\right)
\over  \left( 1-w +i0^+\right)\left( 1-z-w+i0^+ \right)}. \nn
\label{7.03}
\end{eqnarray}
The integral has a discontinuity for $0 \le w \le 1$. As long as $w \not = 1$, the discontinuity is
\begin{eqnarray}
\text{Disc}\, I_1&=& {g^2 C_F\over  8\pi^2 \epsilon}{1 \over 1 -w } \int_0^{1} {\rm d}z
 \left(1-z\right) \delta  \left( 1-z-w+i0^+ \right) \nn
&=& {g^2 C_F \over  8\pi^2 \epsilon}{w \over 1 -w }\theta\left( 0 \le w \le 1 \right) .
\label{7.04}
\end{eqnarray}
The singular terms as $w \to 1$ can be evaluated by a method similar to that used for Eqs~(\ref{6.18},\ref{6.21}) to give
\begin{eqnarray}
\text{Disc}\, I_1&=& {g^2 C_F\over  8\pi^2 \epsilon}\left({w \over 1 -w }\right)_+ \theta\left( 0 \le w \le 1 \right).
\label{7.05}
\end{eqnarray}

Figure~\ref{fig:7}(b) is the discontinuity of
\begin{eqnarray}
I_2 &=& {g^2 C_F\over 4 \pi} \int { {\rm d}^d k \over (2 \pi)^d}\frac12 \text{Tr}\ \xslash{p} \gamma^\mu { \xslash p - \xslash{k} \over (p-k)^2+i0^+} \nn
&& {\xslash{n} \over n \cdot (p-k-r)+i0^+} { \xslash p - \xslash{k} \over (p-k)^2+i0^+} \gamma_\mu {1 \over k^2+i0^+}.\nn
\label{7.06}
\end{eqnarray}
The same manipulations as the previous case give
\begin{eqnarray}
I_2&=& i{g^2 C_F\over  16\pi^3 \epsilon}\int_0^{1} {\rm d}z { z
\over 1-z-w+i0^+ },
\label{7.07}
\end{eqnarray}
so that
\begin{eqnarray}
\text{Disc}\, I_2&=& i{g^2 C_F\over  16\pi^3 \epsilon}\int_0^{1} {\rm d}z z ( - 2 \pi i) \delta \left( 1-z-w+i0^+ \right) \nn
&=& {g^2 C_F \over  8\pi^2 \epsilon} (1-w)  \theta\left( 0 \le w \le 1 \right).
\label{7.08}
\end{eqnarray}

Figure~\ref{fig:7}(c) vanishes, since $n^2=0$. Subtracting half the wavefunction graph for each external quark line gives
\begin{eqnarray}
-\frac 1 2 I_w&=& - {g^2 C_F \over  16\pi^2 \epsilon} \delta\left(1-w\right).
\label{7.09}
\end{eqnarray}

The infinite part of the matrix element is the sum of twice Eq.~(\ref{7.05}), Eq.~(\ref{7.08}), and Eq.~(\ref{7.09}),
\begin{eqnarray}
&&{g^2 C_F\over 8 \pi^2 \epsilon}  \theta\left( 0 \le w \le 1 \right) \nn
&&\times \Biggl[2 \left({w \over 1-w} \right)_+  + 1 - w 
 - \frac 12 \delta(1-w) \Biggr]  \nn
&=&{g^2 C_F\over 8 \pi^2 \epsilon} \left[{1+w^2 \over \left(1 - w\right)_+} + \frac32  \delta(1-w) \right]\theta\left( 0 \le w \le 1 \right) \nn
&=&{\alpha_s \over 2 \pi \epsilon}P_{q \leftarrow q}(w) ,
\label{7.10}
\end{eqnarray}
in terms of the standard Altarelli-Parisi splitting kernel
\begin{eqnarray}
P_{q \leftarrow q}(w) &=&C_F \left[{1+w^2 \over \left(1 - w\right)_+} + \frac32  \delta(1-w) \right]  \theta\left( 0 \le w \le 1 \right).\nn
\label{7.11}
\end{eqnarray}

Equation~(\ref{7.10}) gives the operator renormalization equation
\begin{eqnarray}
O_q^{(0)}\left(y p^+\right) &=& \int {{\rm d}w \over  w} \ Z\left({y \over w} \right) O_q\left(w p^+\right),
\label{7.12}
\end{eqnarray}
with
\begin{eqnarray}
Z\left(z\right)=\delta\left(1-z\right) + {g^2 \over 8 \pi^2 \epsilon}P_{q \leftarrow q}\left(z \right).
\label{7.13}
\end{eqnarray}

If one writes $y p^+=k^+$ and $w p^+=\ell^+$, then
\begin{eqnarray}
O_q^{(0)}\left(k^+\right) &=& \int_{-\infty}^\infty {{\rm d}\ell^+ \over \ell^+} \ Z\left({ k^+\over \ell^+} \right) O_q\left(\ell^+\right),
\label{7.14a}
\end{eqnarray}
which makes no reference to the quark momentum $p$ used to compute the renormalization factor. The renormalization is invariant under boosts in the $z$ direction, under which $+$ components of momentum all get rescaled by a common factor, $p^+ \to \lambda p^+$. Eq.~(\ref{7.14a}) is valid for $k^+ >0$. One can derive a similar expression for $k^+ <0$. The two expressions can be combined into
\begin{eqnarray}
O_q^{(0)}\left(k^+\right) 
&=& \int_{-\infty}^\infty {{\rm d}(\ell^+ /k^+) \over \ell^+/k^+ } \ Z\left({ k^+\over \ell^+} \right) O_q\left(\ell^+\right)\nn
&=& \int_{-\infty}^\infty {{\rm d}\ell^+ \over \abs{ \ell^+} } \ Z\left({ k^+\over \ell^+} \right) O_q\left(\ell^+\right) ,
\label{7.14}
\end{eqnarray}
which is valid for either sign of $k^+$.

Differentiating Eq.~(\ref{7.14}) with respect to $\mu$ gives the renormalization group equation for the quark distribution operator
\begin{eqnarray}
\mu { {\rm d} \over {\rm d} \mu}O_q\left(k^+\right) &=& {\alpha_s \over \pi} \int_{-\infty}^\infty {{\rm d}(\ell^+ /k^+) \over \ell^+/k^+ } \ P_{q \leftarrow q} \left({ k^+\over \ell^+} \right) O_q\left(\ell^+\right). \nn
\label{7.15}
\end{eqnarray}
Taking moments (see Appendix~\ref{app:mom} for the definitions) gives
\begin{eqnarray}
\mu {{\rm d}  \over {\rm d} \mu}  M_N^{(\pm,\infty)} \left[O_q\left(k^+\right)\right] 
&=&  -\gamma_{2,N} M_N^{(\pm,\infty)} \left[O_q\left(k^+\right)\right] ,\nn
\label{7.16}
\end{eqnarray}
where
\begin{eqnarray}
\gamma_{2,N}
&=&  -{\alpha_s(\mu) \over  \pi } M_N \left[P_{q \leftarrow q}\left(z \right) \right]\nn
&=& {\alpha_s(\mu)  \over 2 \pi}C_F \left[ 4 \sum_{j=2}^N {1 \over j} -{2 \over N(N+1)} +1 \right] ,
\label{7.17}
\end{eqnarray}
is the anomalous dimension. For large values of $N$,
\begin{eqnarray}
\gamma_{2,N}
&\to& {\alpha_s(\mu)  \over 2 \pi} C_F \left[ 4 \ln \bN -3  \right] .
\label{7.17a}
\end{eqnarray}

The infinite moments $M_N^{\infty} \left[O_q\left(k^+\right)\right] $ are  local twist-two quark operators $\bar \psi_u \xslash{n} \left(in \cdot \darr{D}\right)^{N-1} \psi_u/2^N$, so the target matrix elements of $M_N^{\infty} \left[O_q\left(k^+\right)\right] $ give the familiar moment sum rules for deep inelastic scattering. The moments $M_N^\infty$ are defined by integrating $k^+$ over $[-\infty,\infty]$, whereas the moments of quark distribution functions are over $0 \le x \le 1$, i.e.~$0 \le k^+ \le P^+$. The matrix element Eq.~(\ref{6.05}) vanishes for $\abs{x} > 1$, and its value for negative values of $x$ is related to the antiquark distribution,
\begin{eqnarray}
f_{q/T}(-x) &=& - f_{\bar q/T} (x)\ .
\label{7.18}
\end{eqnarray}
Thus the matrix elements of $M_N^\infty$ for even $n$, which are the target matrix element of local twist-two operators, are equal to the even moments of the structure function, which sums over quarks and antiquarks. The matrix elements of $M_N^\infty$ for odd $n$ vanish, and do not imply any sum rule for the odd moments of the structure function. The anomalous dimensions of the local twist two operators agree with the moments of the Altarelli-Parisi kernel, Eq.~(\ref{7.17}).

In dimensional regularization, the finite parts of the one-loop quark distribution operator matrix element vanish on-shell, since $\mathbf{p_\perp}=0$, $p^-=0$, and the loop integrals are scaleless. The matrix element is therefore given by its tree-level value $\delta(1-x)$. The matrix elements of the twist-two local operators also vanish on-shell, for the same reason. This also shows that the quark distribution operator is equivalent to its moments (this is not true for the $B$ decay shape function~\cite{Bauer}). The matrix element for off-shell quarks contains logarithms of $p^2/\mu^2$, so the structure function should be evolved down to some hadronic scale before taking the target matrix element so that there are no large logarithms in the matrix element. An important difference from the $B \to X_s \gamma$ shape function~\cite{BFL,Bauer} is that there is no additional matching that has to be performed at the scale $Q^2 \lambda^2$, i.e. $Q^2/\bN^2$.

The moment analysis for deep inelastic scattering has been discussed in detail, even though it may be familiar to many readers. The reason is that there are important differences between the structure functions in deep inelastic scattering and the shape function in the decay of heavy mesons, having to do with the range of the $k^+$ integration in the moment of the quark distribution operator. The results for $B$ decays will be presented elsewhere~\cite{Bauer}.

\subsection{Breit Frame}

Structure function evolution in the Breit frame is given by the graphs in Fig.~\ref{fig:6}(a--d,h--j). Many graphs vanish using $n^2=\bn^2=0$. The only non-zero graphs are Fig.~\ref{fig:6}(b,h--j).

The ultrasoft graph, Fig.~\ref{fig:6}(b), appears to have the form of the ultrasoft vertex correction Eq.~(\ref{5.03}) with an additional propagator for the intermediate $\xi_n$ propagator. There is one very important difference, however. At scales below $Q^2(1-x)$, the momentum $p_2^+$ of the intermediate quark is of order $Q \lambda$, whereas the momentum of the ultrasoft gluon is of order $Q \lambda^2$. Thus the power counting rules of the effective theory imply that the $\xi_n$ propagator denominator in Eq.~(\ref{5.03}), $1/[n \cdot (p_2-k)]$ must be expanded in a power series in $k$, the momentum space analog of the multipole expansion.\footnote{This is analogous to the multipole expansion for the ultrasoft fields in NRQCD~\cite{GR,LMR}.
The multipole expansion is automatic if one makes an additional field redefinition; see the discussion after Eq.~(\ref{7.24}).}
The resulting integral vanishes, so Fig.~\ref{fig:6}(b) is zero.

The collinear graph  Fig.~\ref{fig:6}(i) gives
\begin{eqnarray}
&& {g^2 C_F \over 4 \pi} \frac 1 2 \text{Tr} \frac 1 2 p^+ \xslash{\bar n} 
\gamma^\nu  {\xslash n \over n \cdot (p-r)+i0^+} \nn
&& \int { {\rm d}^d k \over (2 \pi )^d} 
n^\alpha { 1 \over  n \cdot (-k) } \gamma^\mu {\xslash{\bar n}\  n \cdot (p-k)  \over 2 (p - k )^2+i0^+}
{\xslash{n}\ \bar n_\alpha \over 2}   
 {1 \over k^2+i0^+} ,\nn
\label{17.19}
\end{eqnarray}
and Fig.~\ref{fig:6}(j) gives
\begin{eqnarray}
&&{g^2 C_F \over 4 \pi}  \frac 1 2 \text{Tr} \frac 1 2 p^+ \xslash{\bar n} 
\gamma^\nu  {\xslash n  \over n \cdot (p-r-k)+i0^+} \nn
&& \int { {\rm d}^d k \over (2 \pi )^d} 
n^\alpha { 1 \over  n \cdot k } \gamma^\mu {\xslash{\bar n}\  n \cdot (p-k)  \over 2 (p - k )^2+i0^+}
{\xslash{n}\ \bar n_\alpha \over 2}   
 {1 \over k^2+i0^+}, \nn
\label{7.20}
\end{eqnarray}
where the relative minus sign in $\bar n \cdot k$ is because the vertex is from $W^\dagger$ rather than $W$. The sum of the two graphs has
\begin{eqnarray}
&&{1 \over n \cdot (p-r)+i0^+} - {1 \over n\cdot (p-r-k)+i0^+}\nn
&=& - { n \cdot k \over
\left[ n \cdot (p-r)+i0^+\right] \left[ n\cdot (p-r-k) +i0^+\right] }, 
\label{7.21}
\end{eqnarray}
and gives the integral
\begin{eqnarray}
&&4  g^2 C_F T^{\mu \nu}  p^+  {1 \over n \cdot (p-r)+i0^+} \nn
&& \int { {\rm d}^d k \over (2 \pi )^d}  { 1 \over  n \cdot (p-k-r)+i0^+ }  {  n \cdot (p-k)  \over  (p - k )^2+i0^+}  {1 \over k^2+i0^+}, \nn
\label{7.22}
\end{eqnarray}
which is the same as $I_1$ in Eq.~(\ref{7.02}), and so gives Eq.~(\ref{7.05}) for the running of the structure function.

There is no analog of graph Fig.~\ref{fig:7}(b), so its contribution Eq.~(\ref{7.08}) is missing. It is proportional to $1-w\sim\lambda$, and so is of the same order as higher order terms in the power counting which we have dropped.
 
The wavefunction contribution from the collinear graph is the same as the full theory wavefunction contribution, and gives Eq.~(\ref{7.09}). The structure function evolution kernel in the Breit frame is
\begin{eqnarray}
P_{q \leftarrow q}(w) - C_F \left(1-w\right)
\label{7.23}
\end{eqnarray}
The moments of $(1-w)$ vanish as $1/N^2$ for large $N$. One can use $P_{q \leftarrow q}(w)$ for the evolution kernel in Breit frame, since the difference from Eq.~(\ref{7.23}) is higher order in $\lambda$.\footnote{Consistent power counting in the Breit frame automatically implies that $1 - x \sim \lambda \ll 1$, so the Breit frame anomalous dimension Eq.~(\ref{7.23}) is not valid away from $x \to 1$. The target rest frame, however, can also be used for $x$ far from $1$.}

The entire running of the structure function in the Breit frame is from collinear graphs. The ultrasoft graphs vanish. This same result has also been obtained in studying the renormalization of event shape variables using SCET~\cite{event}.

In SCET, one usually makes an additional field redefinition
\begin{eqnarray}
\xi_{\bar n}(x) &=& Y_{\bar n}(x,-\infty) \xi_{\bar n}^{(0)}(x)
\label{7.24}
\end{eqnarray}
where $Y_{\bn}$ is an ultrasoft Wilson line. The new $\bn$-collinear fields $ \xi_{\bar n}^{(0)}$ no longer interact with ultrasoft gluons. In terms of these fields, Eq.~(\ref{6.28a}) becomes
\begin{eqnarray}
O_q(k^+) &=& {1 \over 4 \pi} \int_{-\infty}^\infty \diff z e^{-i z k^+}
\left[ \bar \xi_{\bar n}^{(0)} W_{\bar n}\right]\!\! (n z) \nn
&& Y_{\bar n}(nz,-\infty)^\dagger Y_n(z,0)\, \xslash n
Y_{\bar n}(0,-\infty) \left[ W^\dagger_{\bar n}  \xi_{\bar n}^{(0)}\right]\!\! \left( 0 \right). \nn
\label{7.25}
\end{eqnarray}
After the field redefinition, the collinear graphs are the same as before. The ultrasoft graphs have propagators which are independent of the collinear quark momentum $p$, since the collinear quark momentum $p$ does not enter into the Wilson line. This automatically enforces the multipole expansion. The ultrasoft graphs vanish, since they are scaleless and the entire running of the structure function is from the collinear graphs. This is precisely what we found above.

\section{Deep Inelastic Scattering at one loop \label{sec:dis}}

The deep inelastic scattering structure function to one-loop has been computed in Ref~\cite{Bardeen}. The moments of the non-singlet structure function $F_2/(2x)$ are
\begin{eqnarray}
M_N &=& \left[1 + {\alpha_s \over 4 \pi} B_{2,N}^{\text{NS}} + \gamma_N \ln{\mu \over Q}\right]A_N(\mu),
\label{8.01}
\end{eqnarray}
where $A_N(\mu)$ are the matrix elements of the twist-two operators renormalized at $\mu$.
The anomalous dimension $\gamma_N$ is equal to $\gamma_{2,N}$ in Eq.~(\ref{7.17}),
and
\begin{eqnarray}
B_{2,N}^{\text{NS}} &=& C_F \Biggl[ 3 \sum_{j=1}^N {1 \over j} -
4 \sum_{j=1}^N {1 \over j^2}\nn
&& -{2 \over N(N+1)} \sum_{j=1}^N {1 \over j}
+ 4 \sum_{s=1}^N {1 \over s}\sum_{j=1}^s {1 \over j}\nn
&&+{3 \over N}
+ {4 \over N+1} + {2 \over N^2} - 9 \Biggr] ,
\label{8.03}
\end{eqnarray}
in the $\overline{\text{MS}}$ scheme. For $N \to \infty$,
\begin{eqnarray}
B_{2,N}^{\text{NS}} &\to& C_F \Biggl[2\ln^2 \bN + 3 \ln \bN - { \pi^2 \over 3}  - 9 \Biggr].
\label{8.05}
\end{eqnarray}
Moments of the longitudinal structure function $F_L$ vanish as $1/N$ as $N \to \infty$.

These results can be compared with the calculations in this paper. The tensor structure in Eq.~(\ref{6.10}) shows that $F_L=0$.
The  result for $F_1=F_2/(2x)$ is the product of the square of the matching coefficient at $Q$, Eq.~(\ref{4.10}), twice the running from $Q^2$ to $Q^2/\bN$ using Eq.~(\ref{5.12}), the matching coefficient at $Q^2/\bN$, Eq.~(\ref{6.27}), and the running from $Q^2/\bn$ to $\mu$ using Eq.~(\ref{7.17}):
\begin{eqnarray}
{M_N \over A_N(\mu)}  &=& C^2(Q) e^{2 \gamma_1 \ln{ [(Q/\sqrt{\bN})/Q]}}\nn
&& \times \left[1 + M_N(\mathcal{M})\right]  e^{ \gamma_2 \ln{  [\mu/(Q/\sqrt{\bN}}]} +
\mathcal{O}\left(\alpha_s^2\right)\nn
&=&1 + C_F {\alpha_s \over 2 \pi}  \left[   -8 +{\pi^2 \over 6} \right]\nn
&&- C_F {\alpha_s \over 2 \pi} \left[ \ln^2 {{Q^2 /\bN} \over Q^2} + 3 \ln{{Q^2 /\bN} \over Q^2} \right]\nn
&&+{\alpha_s \over 2 \pi }C_F \left[ \frac 7 2  - {\pi^2 \over 3} \right] \nn
&& +  \frac 1 2 \gamma_N \ln{ \mu^2 \over Q^2/\bN } + \mathcal{O}\left(\alpha_s^2\right)\nn
&=& 1 + C_F {\alpha_s \over 4 \pi} \Bigl[-9 - {\pi^2/3}  -  2 \ln^2 \bN + 6 \ln \bN \Bigr]\nn
&& +  \gamma_N \ln{\mu \over Q} + \frac 1 2 \gamma_N \ln \bN \ .
\label{8.06}
\end{eqnarray}
Using Eq.~(\ref{7.17a}) for the anomalous dimension, Eq.~(\ref{8.06}) agrees with Eqs.~(\ref{8.01},\ref{8.05}).
 
\section{SCET Anomalous Dimensions \label{sec:gamma}}

There is a non-trivial relation between the anomalous dimensions and the matching conditions in the effective theory. The anomalous dimension for scaling the current between $Q^2$ and $Q^2/\bN $ has the form
\begin{eqnarray}
\gamma_1(\mu) &=& A\!\left(\alpha_s(\mu),\ln{\mu \over Q} \right) .
\label{9.01}
\end{eqnarray}
The anomalous dimension can depend on $Q$ since that is a label on the SCET fields. It does not depend on $\bar N$, since at this stage, a single current is being renormalized, which has no information about $1-x$. The anomalous dimension for scaling the structure function between $Q^2/\bN$ and $\mu^2$ has the form
\begin{eqnarray}
\gamma_2(\mu) &=& B\!\left(\alpha_s(\mu),\ln \bN \right) ,
\label{9.02}
\end{eqnarray}
where the second argument has been chosen to be $\ln \bN$ rather than $\bN$ for later convenience. The anomalous dimension can no longer depend on $Q$, since it has been integrated out of the effective theory. It can depend on $\bN$, since the parton distribution operators know about $1-x$. At the matching scale between the two regimes, one has the matching coefficient for the structure function
\begin{eqnarray}
C\!\left(\alpha_s(\mu), \ln {\bN \mu^2 \over Q^2} \right)
\label{9.03}
\end{eqnarray}
The matching coefficient can only depend on the relevant mass scale at that point, which is the invariant mass of the hadronic jet, $Q^2/\bN$.

The final answer for the structure function is given by taking the square of the matching coefficient at $Q$, running it using twice Eq.~(\ref{9.01}) to $\mu^2 \sim Q^2/\bN$, multiplying by $C$  (note that the matching factor is multiplicative),  scaling by Eq.~(\ref{9.02}) to a low scale, and then taking the target matrix element. The result must be independent of the choice of matching scale $\mu$. This gives
\begin{eqnarray}
{1\over C}\ \mu { {\rm d} \over {\rm d} \mu} C\left(\alpha_s(\mu), \ln {\bN \mu^2 \over Q^2} \right) &=& \gamma_2(\mu) - 2 \gamma_1(\mu).\nn
\label{9.04}
\end{eqnarray}
Let $D_{1,2}C$ be the derivatives of $C$ with respect to the first and second argument. Equation~(\ref{9.04}) gives
\begin{eqnarray}
\left( \beta(\alpha_s) D_1 + 2 D_2 \right) \ln C&=&\gamma_2 - 2 \gamma_1.
\label{9.05}
\end{eqnarray}
where
\begin{eqnarray}
\mu { {\rm d} \over {\rm d} \mu} \alpha_s(\mu) &=& \beta(\alpha_s).
\label{9.05a}
\end{eqnarray}

Letting $\left( \beta(\alpha_s) D_1 + 2 D_2 \right)\ln C \equiv F$, one gets
\begin{eqnarray}
F\left(\alpha_s, \ln {\bN } + \ln { \mu^2 \over Q^2} \right) &=&B\left(\alpha_s,\ln \bN \right) - 2 A\left(\alpha_s,\ln{\mu \over Q} \right) .\nn
\label{9.06}
\end{eqnarray}
Equation~(\ref{9.06}) implies that $A$, $B$ and $F$ are at most linear in their second arguments to all orders in $\alpha_s$. This has been shown previously~\cite{KorRad} by analyzing the momentum integrals of the Feynman graphs. Letting
\begin{eqnarray}
\gamma_1 &=& A_1 \left(\alpha_s \right)\ln{\mu^2 \over Q^2} +A_0 \left(\alpha_s \right), \nn
\gamma_2 &=& B_1 \left(\alpha_s \right)\ln{\bN } +B_0 \left(\alpha_s \right) , \nn
F &=&F_1 \left(\alpha_s \right)\ln {\mu^2 \bN \over Q^2} +F_0 \left(\alpha_s \right), \nn
\ln C &=&  \sum_{r=0}^\infty C_r \left(\alpha_s \right)\ln^r {\mu^2 \bN \over Q^2} ,
\label{9.07}
\end{eqnarray}
gives
\begin{eqnarray}
2 A_1\left(\alpha_s \right) &=& - B_1 \left(\alpha_s \right) =
 - F_1 \left(\alpha_s \right) ,\nn
F_0 \left(\alpha_s \right) &=& B_0 \left(\alpha_s \right) - 2 A_0\left(\alpha_s \right),\nn
\beta(\alpha_s) C_1^\prime \left(\alpha_s \right) + 4  C_{2} \left(\alpha_s \right) &=& F_1, \nn
\beta(\alpha_s) C_0^\prime \left(\alpha_s \right) + 2  C_{1} \left(\alpha_s \right) &=& F_0 . 
\label{9.08}
\end{eqnarray}
and
\begin{eqnarray}
\beta(\alpha_s) C_r^\prime \left(\alpha_s \right) + 2(r+1)  C_{r+1} \left(\alpha_s \right) &=& 0
\label{9.08a}
\end{eqnarray}
if $r > 1$, where the prime denotes a derivative with respect to $\alpha_s$.

The conditions on $C_i$ are the usual relations that the logarithms in the matching condition are the difference of the anomalous dimensions of the theories on either side. Here, however, we have the additional constraint $2A_1=-B_1$ relating the two anomalous dimensions. Equations~(\ref{5.12},\ref{6.27},\ref{7.17a}) give for the one-loop values
\begin{eqnarray}
A_0\left(\alpha_s \right) &=& - {3 \alpha_s \over 2 \pi} C_F, \nn
A_1\left(\alpha_s \right) &=& - {\alpha_s \over \pi} C_F ,\nn
B_0\left(\alpha_s \right)  &=& -{3 \alpha_s \over 2\pi} C_F, \nn
B_1\left(\alpha_s \right)  &=& {2 \alpha_s \over \pi} C_F ,\nn
C_2\left(\alpha_s \right)  &=& {\alpha_s \over 2 \pi} C_F, \nn
C_1\left(\alpha_s \right)  &=& {3 \alpha_s \over 4 \pi} C_F ,
\label{9.10}
\end{eqnarray}
and $A_r=B_r=0$, $r>1$ and $C_r=0$, $r>2$, which satisfy Eq.~(\ref{9.08}), since $\beta(\alpha_s)$ starts at order $\alpha_s^2$.

\section{Exponentiation \label{sec:exp}}

The final result for the moments of the structure function computed using SCET is 
\begin{eqnarray}
F_N(Q^2) &=&  C^2(Q) e^{-I_1} \left[1 + M_N(\mathcal{M})\right] e^{-I_2}A_N(\mu_0),\nn
\label{10.01}
\end{eqnarray}
with
\begin{eqnarray}
I_1 &=& \int_{Q/\sqrt{\bN}}^Q {  {\rm d \mu} \over \mu}\ 2 \gamma_1(\mu),\nn
I_2 &=& \int^{Q/\sqrt{\bN}}_{\mu_0} {  {\rm d \mu} \over \mu} \ \gamma_2(\mu),
\label{10.02}
\end{eqnarray}
where $\mu_0$ is some reference scale of order a few GeV, and $C(Q)$ $\mathcal{M}$, $\gamma_1$ and $\gamma_2$ are given in Eqs.~(\ref{4.10},\ref{6.28},\ref{5.12},\ref{7.17a}), and $A_N$ is the target matrix element of the local twist-two operator.  All the large logarithms are contained in the exponent $I=I_1+I_2$. Comparing with Eq.~(\ref{1.02}), we see that $f_0$ is obtained by integrating the anomalous dimension using the one-loop values for $A_1$ and $B_1$, $f_1$ is obtained by integrating the two-loop values for $A_1,B_1$ and the one-loop values for $A_0,B_0$, etc.

In deriving Eq.~(\ref{10.01}), we have assumed that $Q^2 > Q^2/\bN > \mu_0^2$. For fixed $Q^2$ and $\bN \agt Q^2/\mu_0^2$, this inequality no longer holds.  The effective theory result for such large moments can be computed as follows. Match the current at $Q$ and evolve using $\gamma_1$ from $Q$ to $\mu_0$. At the scale $\mu_0$, one computes the matrix element of the time-ordered product of two currents in the target hadron. This is a non-perturbative computation since $\mu_0 \sim \lqcd$ is a hadronic scale. For moments with  $\bN \sim Q^2/\mu_0^2$, the scattering process is in the resonance region since $p_X^2 \sim \mu_0^2 \sim \lqcd^2$, and is described by exclusive form-factors rather than the deep inelastic structure function. In equations, the result is
\begin{eqnarray}
F_N(Q^2) &=&  C^2(Q) e^{-I_1^\prime} A_N^\prime(\mu_0),\nn
\label{10.01a}
\end{eqnarray}
where
\begin{eqnarray}
I_1^\prime &=& \int_{\mu_0}^Q {  {\rm d \mu} \over \mu}\ 2 \gamma_1(\mu),\nn
\label{10.02a}
\end{eqnarray}
and $A_N^\prime(\mu_0)$ is the $N^{\text{th}}$ moment of the matrix element of the time-ordered product of two SCET currents at the scale $\mu_0$.

The computation of moments is not uniformly convergent over the parameter range. One can formally take the limit $Q^2 \to \infty$ first, and then study arbitrarily large moments, since $Q^2/\bN$ is now always much greater than $\mu_0^2$. However, 
if one takes $Q^2$ large but finite, then for large enough $\bN$, $Q^2/\bN$ can become smaller than $\mu_0^2$. The relevant limit for fixed target experiments is to keep the beam energy fixed, in which case the maximum value of $Q^2$ is $2 M_T E x$, and depends on $x$.

\subsection{Landau Pole\protect\footnote{I would like to thank I.Z.~Rothstein for discussions on this section}}

The anomalous dimension integrals in Eq.~(\ref{10.01}) contain $\alpha_s(\mu)$ where $\mu$ varies between $Q$ and $\mu_0$. As discussed in the preceeding paragraph, this is true even for $\bN \to \infty$ where formula Eq.~(\ref{10.01}) must be modified. The SCET computation therefore does not suffer from any Landau pole singularities~\cite{Landau}. For large enough moments, or equivalently, for $1-x \sim \lqcd^2/Q^2$, the scattering cross-section depends on resonance physics; this is a true non-perturbative contribution, not a breakdown of the perturbation series.

The relation $2A_1=-B_1$ allows us to write the renormalization group integration as a double integral~\cite{Catani}. It is easier to  derive the result starting from the answer. Consider the integral
\begin{eqnarray}
I_c &=&\int_{1/\bN }^1 {{\rm d} y \over y} \Biggl[
\int_{ \mu_0 }^{\sqrt y Q}  2 \Gamma_1 \left( \alpha_s \left(\mu \right)\right)  { {\rm d}\mu \over \mu}
+\Gamma_2 \left( \alpha_s \left( \sqrt y Q \right) \right)\Biggr] . \nn
\label{9.11}
\end{eqnarray}
Changing the order of the integrals gives
\begin{eqnarray}
I_c&=& \int_{Q /\sqrt{\bN} }^Q  { {\rm d}\mu \over \mu}  \Bigl[
  2 \Gamma_1\left( \alpha_s \left(\mu \right)\right)\ln { Q^2 \over \mu^2  }+2  \Gamma_2 \left( \alpha_s \left(\mu \right)\right)  \Bigr] \nn
&& +\int_{\mu_0 }^{Q/\sqrt{\bN} }  { {\rm d}\mu \over \mu}  \left[
 2\Gamma_1 \left( \alpha_s \left(\mu \right)\right)\ln {\bN } \right]  ,
\label{9.12}
\end{eqnarray}
which has the same form as the integration of anomalous dimensions from $Q \to Q/\sqrt{\bN}$ and from $Q/\sqrt{\bN} \to \mu_0$. 
Let
\begin{eqnarray}
I_1 + I_2 &=& I_c + \int_{\mu_0 }^{Q} { {\rm d}\mu \over \mu}  
  \Gamma_3 \left( \alpha_s \left(\mu \right)\right). 
\label{9.12a}
\end{eqnarray}
Comparing with Eqs.~(\ref{9.07},\ref{9.10},\ref{10.02}) gives
\begin{eqnarray}
2\gamma_1 &=& 2 \Gamma_1 \ln {Q^2\over \mu^2} + 2 \Gamma_2 + \Gamma_3 ,\nn
\gamma_2 &=& 2\Gamma_1 \ln \bN + \Gamma_3 ,
\label{9.12c}
\end{eqnarray}
so that
\begin{eqnarray}
\Gamma_1 &=& - A_1 = \frac 1 2 B_1 ={\alpha_s \over \pi} C_F ,\nn
\Gamma_2 &=& A_0 - \frac 1 2 B_0  = -{3\alpha_s \over 4\pi} C_F, \nn
\Gamma_3 &=&  B_0= -{3 \alpha_s \over 2 \pi} C_F ,
\label{9.13}
\end{eqnarray}
where the last entry on each line is the one-loop value.  Eq.~(\ref{9.11}) can also be written as
\begin{eqnarray}
I_c &=&\int_{0 }^1 {\rm d} z  {z^{N-1}-1\over 1-z} \Biggl[
\int_{ \mu_0^2/(1-z) }^{Q^2}  \Gamma_1 \left( \alpha_s \left((1-z) k^2 \right)\right)  { {\rm d}k^2 \over k^2} \nn
&& +\Gamma_2 \left( \alpha_s \left( (1-z) Q^2 \right) \right)\Biggr] . 
\label{9.11a}
\end{eqnarray}
The equality between Eq.~(\ref{9.11}) and Eq.~(\ref{9.11a}) is derived in Sec.~5 of Ref.~\cite{Catani}. The usual expressions for the resummed structure function derived using factorization methods have the form Eq.~(\ref{9.11}) with $\mu_0 \to \mu_0 \sqrt y$ or Eq.~(\ref{9.11a}) with $\mu_0^2/(1-z) \to \mu_0^2$~\cite{Catani}, and both suffer from Landau pole singularities~\cite{Landau}, as does the ratio of the moments of the Drell-Yan cross-section to the square of the moments of the deep inelastic cross-section, $\Delta_N$, which is independent of the hadron structure function~\cite{Catani},
\begin{eqnarray}
&& - \frac 1 2 \ln \Delta_N \nn
&=&  \int_{0 }^1 {\rm d} z  {z^{N-1}-1\over 1-z} \Biggl[
\int_{ Q^2 (1-z) }^{Q^2}  \Gamma_1 \left( \alpha_s \left((1-z) k^2 \right)\right)  { {\rm d}k^2 \over k^2} \nn
&& +\Gamma_2 \left( \alpha_s \left( (1-z) Q^2 \right) \right)\Biggr] . 
\label{9.11b}
\end{eqnarray}
However, the corresponding SCET computation does not have any Landau pole singularities. The Landau pole singularity  arises on converting the SCET anomalous dimension integration to the form Eq.~(\ref{9.11}).  The conversion is only valid for $Q^2/\bN > \mu_0^2$ in deep inelastic scattering, and $Q^2/\bN^2 > \mu_0^2$ in Drell-Yan.\footnote{The extra $\bN$ for Drell-Yan arises because of the difference in kinematics.} The factorization form, Eq.~(\ref{9.11a},\ref{9.11b}) introduces a spurious Landau pole singularity, because it is equivalent to using Eq.~(\ref{10.01}) for large moments, where it is no longer valid, rather than replacing it by the correct expression, Eq.~(\ref{10.01a}), discussed earlier.

The SCET form for $\Delta_N$ does not have a Landau pole singularity, but instead contains a non-perturbative resonance contribution. The resonance contributions enter $\Delta_N$ for large moments even though $\Delta_N$ does not depend on the structure function, because non-perturbative effects enter Drell-Yan at $\bN \sim Q/\mu_0$ before they enter deep inelastic scattering at $\bN \sim Q^2/\mu_0^2$, so they appear in $\Delta_N$ beginning at $\bN \sim Q/\mu_0$.

The SCET form for the shape function in $B$ decays also does not have a Landau pole singularity.

\subsection{Cusp Anomalous Dimension}

\begin{figure}
\includegraphics[width=3cm]{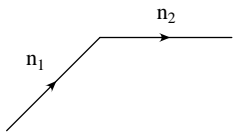}
\caption{Wilson line along the $n_1$ and $n_2$ directions. \label{fig:cusp}}
\end{figure}

$\Gamma_1$ is known as the cusp anomalous dimension, because it can be computed from the anomalous dimension of a Wilson line with a cusp~\cite{KorRad} (see Fig.~\ref{fig:cusp}).
The anomalous dimension of the Wilson line in Fig.~\ref{fig:cusp} has the form
\begin{eqnarray}
\gamma_W\!\! \left( \alpha_s, \eta\right), \qquad \eta^2 = {\left(2 n_1 \cdot n_2\right)^2 \over n_1^2 n_2^2 },
\label{9.21}
\end{eqnarray}
since it must be Lorentz invariant, and invariant under rescaling of the vectors $n_i$. As $\eta \to \infty$,
\begin{eqnarray}
\gamma_W \left( \alpha_s, \eta \right)  &\to& \ln \eta\ \Gamma_c\left( \alpha_s \right),
\label{9.15}
\end{eqnarray}
which defines the cusp anomalous dimension $\Gamma_c$~\cite{KorRad}. To two loop order~\cite{KorRad}
\begin{eqnarray}
\Gamma_c &=& {\alpha_s \over \pi} C_F + \left( {\alpha_s \over \pi}  \right)^2
C_F \left[ C_A \left( \frac{67}{36}-\frac {\pi^2} {12}  \right) - \frac 5 {9} T_F N_F \right].\nn
\label{9.16}
\end{eqnarray}

We have seen earlier that the equality of the anomalous dimensions in the target rest frame and Breit frame implies that the $\ln \mu$ terms in the anomalous dimension from the ultrasoft and collinear graphs are related, so that $A_1=-\Gamma_1$ can be determined from the ultrasoft graph alone. The anomalous dimension Eq.~(\ref{9.21}) is computed with two Wilson lines with $n_i^2 \not =0$. The ultrasoft anomalous dimension computed earlier is obtained by taking the limit $n_1 \to n$, $n_2 \to \bn$, which are both null vectors. Using $p_1 \propto n_1$ and $p_2 \propto n_2$, one has
\begin{eqnarray}
\eta^2=  {\left(2 n_1 \cdot n_2\right)^2 \over n_1^2 n_2^2 } &=& { Q^4 \over p_1^2 p_2^2} ,
 \label{9.17}
\end{eqnarray}
and
\begin{eqnarray}
\ln \eta &=& \frac12 \ln{ Q^4 \over p_1^2 p_2^2} .
 \label{9.18}
\end{eqnarray}
The graph with $p_i^2 \to 0$ in pure dimensional regularization replaces $p_i^2$ by $\mu^2$, so that
\begin{eqnarray}
\ln \eta &=&   \ln{ Q^2 \over \mu^2} .
 \label{9.19}
\end{eqnarray}
Comparing Eq.~(\ref{9.15},\ref{9.19}) with Eq.~(\ref{9.07}) gives
\begin{eqnarray}
\Gamma_c \left( \alpha_s\right) =  -A_1 \left( \alpha_s\right)= \Gamma_1 \left( \alpha_s\right) =
\frac 1 2 B_1 \left( \alpha_s\right) .
 \label{9.20}
\end{eqnarray}
Using Eq.~(\ref{9.20}) for $A_1$, $B_1$, and Eq.~(\ref{9.10}) for $A_0$, $B_0$ in Eq.~(\ref{10.01}) gives $F_N$ including all terms of order $\alpha_s$ that do not vanish as $N \to \infty$, and including the first two series $f_{0,1}$ in Eq.~(\ref{1.02}). 

\section{Conclusions \label{sec:conc}}

The deep inelastic structure function was computed in the $x \to 1 $ endpoint region using SCET. The effective theory properly separates the scale $Q^2$, $Q^2(1-x)$ and $\lqcd^2$, and allows the problem to be analyzed one scale at a time. The structure function is equivalent to local twist-two operators below the scale $Q^2(1-x)$. The scale $Q^2(1-x)^2$ does not play a special role in the analysis of structure functions. This is very different from the case of $B \to X_s \gamma$ decay, where the shape function has an anomalous dimension between $Q^2(1-x)$ and $Q^2(1-x)^2$, and stops running below $Q^2(1-x)^2$~\cite{BFL,Bauer}. Consistency of the effective theory implies that the SCET anomalous dimensions are linear in $\ln \mu$ to all orders in perturbation theory. The SCET formulation also avoids the Landau pole singularity in the resummed cross-section.

\begin{acknowledgments}

I would like to thank C.~Bauer, I.Z.~Rothstein, I.W.~Stewart, and M.B.~Wise for discussions. Some of this work was done at the Aspen Center for Physics. This work was supported in part by the Department of Energy under contract DOE-FG03-97ER40546.

\end{acknowledgments}

\appendix

\section{Moments}\label{app:mom}

Let $f(z)$, $g(z)$ and $h(z)$ be functions defined for $0 \le z \le 1$. The convolution of $f$ and $g$, $h=f\star g=g\star f$ is defined by
\begin{eqnarray}
h\left(z\right) &=& \int_0^1 {\rm d}x \int_0^1 {\rm d}y \ \delta(z-xy) f\left(x\right) g\left(y\right) \nn
&=& \int_z^1 { {\rm d}y \over y}  f\left({z \over y}\right) g\left(y\right) \nn
&=& \int_z^1 { {\rm d}x \over x}  f\left(x\right) g\left({z \over x}\right)  .
\label{a.01}
\end{eqnarray}
The moments of a function are defined by
\begin{eqnarray}
M_N\left[ f\left(z\right) \right] &=& \int_0^1 {\rm d}z\ z^{N-1} f(z),
\label{a.02}
\end{eqnarray}
and satisfy the relation
\begin{eqnarray}
M_N\left[ f\star g \right] &=& M_N\left[ f \right] M_N\left[ g\right] .
\label{a.03}
\end{eqnarray}

The moments we need are
\begin{eqnarray}
M_N\left[ \delta(1-z) \right] &=& 1,\nn
M_N\left[ 1 \right] &=& {1 \over N} ,\nn
M_N\left[ 1-z  \right] &=& {1\over N(N+1)} ,\nn
M_N\left[ \left( {z^r \over 1-z} \right)_+  \right] &=& 
- \sum_{j=1+r}^{N-1+r} {1 \over j} \nn
&=& H_r - H_{N-1+r}, \nn
M_N\left[ \left( {\ln (1-z) \over 1-z} \right)_+  \right]
&=& \sum_{j=1}^{N-1} {H_j \over j},
\label{a.04}
\end{eqnarray}
where the harmonic number $H_r$ is defined by
\begin{eqnarray}
H_r &=& \sum_{j=1}^r {1 \over j}.
\label{a.05}
\end{eqnarray}
The large $N$ limits of the moments are ($\bN =N e^{\gamma_E}$)~\cite{Catani}
\begin{eqnarray}
M_N\left[ \delta(1-z) \right] &\to& 1,\nn
M_N\left[ 1  \right] &\to& 0 ,\nn
M_N\left[ 1-z  \right] &\to& 0, \nn
M_N\left[ \left( {z^r \over 1-z} \right)_+  \right] &\to&  H_r - \ln \bN ,\nn
M_N\left[ \left( {\ln (1-z) \over 1-z} \right)_+  \right] &\to&  \frac 1 2 \ln^2 \bN +  \frac 1 2 \zeta(2).
\label{a.06}
\end{eqnarray}

For functions defined on $(-\infty,\infty)$, one can define the moments over $[0,1]$ as in Eq.~(\ref{a.02}) as well as the half-infinite and infinite moments
\begin{eqnarray}
M_N^+\left[ f\left(z\right) \right] &=& \int_0^\infty {\rm d}z\ z^{N-1} f(z) ,\nn
M_N^-\left[ f\left(z\right) \right] &=& \int_{-\infty}^0 {\rm d}z\ z^{N-1} f(z) ,\nn
M_N^\infty\left[ f\left(z\right) \right] &=& \int_{-\infty}^\infty {\rm d}z\ z^{N-1} f(z) \nn
&=&M_N^-\left[ f\left(z\right) \right]  + M_N^+\left[ f\left(z\right) \right].
\label{a.07}
\end{eqnarray}

\end{document}